\documentclass[12pt,a4paper,DIV12]{scrartcl}
\usepackage[utf8]{inputenc}     
\usepackage[T1]{fontenc}
\usepackage{lmodern}
\usepackage[british]{babel}
\usepackage{amsmath}
\usepackage{amssymb}
\usepackage{amsfonts}
\usepackage{color}
\usepackage{float}
\usepackage{appendix}
\usepackage[pdftex]{graphicx}
\usepackage{hyperref}
\usepackage{textcomp}
\usepackage{subfigure}

\newcommand{\I}{\ensuremath{\mathrm{i}\hspace{1pt}}}

\newcommand{\tr}{\ensuremath{\mathrm{tr}}}
\newcommand{\pf}{\ensuremath{\mathrm{Pf}}}
\newcommand{\re}{\ensuremath{\mathrm{Re}}}
\newcommand{\dt}{\ensuremath{\mathrm{det}}}
\newcommand{\sign}{\ensuremath{\mathrm{sign}}}
\newcommand{\api}{\text{a--}\pi}
\newcommand{\Nt}{N_\tau}
\newcommand{\Ncon}{N_{\text{conf.}}}
\newcommand{\rne}{r_0^e}
\newcommand{\kcd}{\kappa_c^\textrm{dec}}
\newcommand{\bcd}{\beta_c^\textrm{dec}}
\newcommand{\Eqref}[1]{Eq.~(\ref{#1})}
\newcommand{\arxiv}[1]{arXiv:\,\href{http://arxiv.org/abs/#1}{{\tt #1}}}
\title{
  {\vspace{-20mm}\normalsize
   \hfill\parbox[b][30mm][t]{35mm}{\textmd{MS-TP-14-20}}}\\[-18mm]
Phase structure of the $\mathcal{N}=1$ supersymmetric Yang-Mills theory 
at finite temperature}

\author{G.~Bergner\\
\textit{\large Universit\"at Frankfurt, Institut f\"ur Theoretische Physik}\\
\textit{\large Max-von-Laue-Str.~1, D-60438 Frankfurt am Main, Germany}\\
\textit{\large E-mail: bergner@th.physik.uni-frankfurt.de}\\[5mm]
P.~Giudice, G.~M\"unster, S.~Piemonte, D.~Sandbrink\\
\textit{\large Universit\"at M\"unster, Institut f\"ur Theoretische Physik}\\
\textit{\large Wilhelm-Klemm-Str.~9, D-48149 M\"unster, Germany}\\
\textit{\large E-mail: munsteg, p.giudice, spiemonte, dirk.sandbrink@uni-muenster.de}
\vspace*{5mm}}

\date{May 13, 2014}

\begin{document}

\maketitle

\begin{abstract}
Supersymmetry (SUSY) has been proposed to be a central concept for the
physics beyond the standard model and for a description of the strong
interactions in the context of the AdS/CFT correspondence. A deeper
understanding of these developments requires the knowledge of the properties
of supersymmetric models at finite temperatures. We present a Monte Carlo
investigation of the finite temperature phase diagram of the $\mathcal{N}=1$
supersymmetric Yang-Mills theory (SYM) regularised on a space-time lattice.
The model is in many aspects similar to QCD: quark confinement and fermion
condensation occur in the low temperature regime of both theories. A
comparison to QCD is therefore possible. The simulations show that for
$\mathcal{N}=1$ SYM the deconfinement temperature has a mild dependence on
the fermion mass. The analysis of the chiral condensate susceptibility
supports the possibility that chiral symmetry is restored near the
deconfinement phase transition.
\end{abstract}

\section{Introduction}

Gauge theories at finite temperatures have been explored intensively by
means of Monte Carlo simulations on a lattice. For Yang-Mills theories
without fermions many calculations have been done for different gauge
groups, see for example \cite{Engels:1989fz,Lucini:2002ku}. A phase
transition has been found, separating a low-temperature phase with
confinement of static quarks from a high-temperature deconfined phase. In
full QCD, including also up, down, and strange quarks, a crossover separates
the confined nuclear matter phase at low temperatures from the quark-gluon
plasma in the high temperature regime
\cite{Aoki:2006we,Aoki:2005vt,Bazavov:2009zn}. The realisation of chiral
symmetry in QCD is another temperature dependent phenomenon. At low
temperatures chiral symmetry is broken, while it is restored at high
temperatures. This provides another (pseudo-)critical temperature. Recent
numerical investigations have shown that the restoration of chiral symmetry
takes place near the deconfinement transition \cite{Aoki:2006br}. The
physical relation between the two critical temperatures remains, however,
unclear due to the lack of an exact order parameter \cite{Borsanyi:2010bp}.

For supersymmetric Yang-Mills theory (SYM) there are only few
non-perturbative results about its behaviour at finite temperatures. A great
interest in the subject comes from the application of the AdS/CFT conjecture
\cite{Maldacena:1997re} to the description of the deconfinement transition
of QCD. The AdS/CFT conjecture is a duality between low-energy string theory
in ten dimensions and strong coupling $\mathcal{N} = 4$ SYM in four
dimensions \cite{Gubser:2009md}. $\mathcal{N} = 4$ SYM is a conformal theory
and therefore reductions are needed in order to relate the results to a
theory like QCD with a mass-gap. Finite temperature is a possibility to
break both supersymmetry and conformal invariance of $\mathcal{N} = 4$ SYM
\cite{Gubser:2009md} and therefore it could be possible that many
fundamental properties are shared between the weakly interacting quark-gluon
plasma and supersymmetric models at finite temperature.

Supersymmetric models at finite temperatures have a different behaviour than
other models, due to the difference between the thermal statistic of Bose
and Fermi particles. In the Euclidean time direction periodic and
anti-periodic boundary conditions must be imposed on fermionic and bosonic
fields, respectively. At zero temperature, in the infinite volume limit,
this difference can be neglected and exact SUSY can be formulated
consistently. At finite temperatures, the temporal direction is compactified
and boundary conditions will break the supersymmetry between fermions and
bosons \cite{Girardello:1980vv}. Therefore there is no high temperature
limit in which a possible spontaneously or explicitly broken supersymmetry
can be effectively restored \cite{Clark:1982xg}. This intriguing property
was subject of many studies in the past, in particular for understanding the
nature and the pattern of this temperature induced SUSY breaking, see
\cite{Das:1989cj} for a review.

Supersymmetry opens the possibility to study the relation between the
deconfinement transition and chiral symmetry restoration. Dual gravity
calculations proved that confinement implies chiral symmetry breaking for a
class of supersymmetric Yang-Mills theories
\cite{Gubser:2009md,Aharony:2006da}.

The object of our investigations is $\mathcal{N} = 1$ SYM at finite
temperatures. This theory describes the strong interactions between gluons
and their superpartners, the gluinos, which are Majorana fermions in the
adjoint representation of the gauge group. At zero temperature the theory is
in a confined phase and chiral symmetry is spontaneously broken by a
non-vanishing expectation value of the gluino condensate. This theory has
been subject of intensive theoretical investigations. Relations between SYM
and QCD have been found in terms of the orientifold planar equivalence
\cite{Armoni:2004uu}. They have lead to conjectures about SYM relics in QCD
\cite{Armoni:2003fb}. $\mathcal{N}=1$ SYM has also a crucial role in the
context of the gauge/gravity duality of the $\mathcal{N}=4$ theory
\cite{Girardello:1999bd}. Numerical simulations of $\mathcal{N} = 1$ SYM are
possible with the Monte Carlo methods \cite{Montvay:1995ea} and they provide
an important non-perturbative tool for exploring the phase diagram at finite
temperatures.

A mass term for the gluinos is added in our numerical simulations and the
results are extrapolated to the chiral limit. The gauge group chosen is
SU(2). The results obtained show clearly that deconfinement occurs at a
temperature which decreases with decreasing gluino mass. The distribution of
the order parameter and the finite size scaling support the possibility that
the order of the associated phase transition is the same for a pure gauge
theory and its supersymmetric extension, at least for the range of masses
considered. Possible scenarios for the relation between chiral symmetry
breaking and deconfinement are also discussed. The chiral symmetry is found
to be restored near the temperature of the deconfinement phase transition,
even if it requires a more careful extrapolation to the chiral limit. A
chiral phase transition of the same order of the deconfinement transition is
argued considering the general symmetries of the model.

\section{Supersymmetric Yang-Mills theory}

The $\mathcal{N} = 1$ SYM theory is the supersymmetric extension of pure
gauge theory. The model is constructed imposing SU($N_c$) gauge invariance
and a single conserved supercharge, obeying the algebra
\begin{equation}
 \{ Q_\alpha, Q_\beta\} 
= (\gamma^\mu C)_{\alpha \beta} P_\mu \quad (\alpha, \beta=1,\dots,4), 
\end{equation}
where the generators of the supersymmetry $Q_\alpha$ are Majorana spinors,
$C$ is the charge conjugation matrix and $P_\mu$ the momentum operator. The
theory contains gluons as bosonic particles, and gluinos as their fermionic
superpartners. The gluino is a spin-\textonehalf~Majorana fermion in the
adjoint representation of the gauge group. A Majorana fermion obeys the
``reality'' condition
\begin{equation}
 \bar{\lambda}(x) = (\lambda(x))^T C.
\end{equation}

Supersymmetry relates the gauge fields $A_\mu(x)$ and gluino fields
$\lambda(x)$:
\begin{eqnarray}
A_\mu(x) & \rightarrow & A_\mu(x) -2 \,\I \bar{\lambda}(x)\gamma_\mu \epsilon \\
\lambda^a(x) & \rightarrow & \lambda^a(x) 
- \sigma_{\mu\nu} F^a_{\mu\nu}(x) \epsilon,
\end{eqnarray}
where $\epsilon$ is a global Majorana fermion, parametrising the
transformation.

The Euclidean on-shell action for $\mathcal{N} = 1$ SYM theory in the
continuum is
\begin{equation}
\label{continuum_action}
S(g,m) =  \int d^4 x \left\{\frac{1}{4} (F_{\mu\nu}^a F_{\mu\nu}^a) 
+ \frac{1}{2}\bar{\lambda}_a (\gamma^\mu D^{ab}_\mu + m)\lambda_b - 
\frac{\Theta}{16\pi}\epsilon_{\mu\nu\rho\sigma}F^{\mu\nu}F^{\rho\sigma}\right\}.
\end{equation}
The $\Theta$-term can be added to the action as in QCD without violating the
underlying symmetries of the model. The operator
$\epsilon_{\mu\nu\rho\sigma}F^{\mu\nu}F^{\rho\sigma}$ is topologically
invariant and the theory is periodic in the parameter $\Theta$, i.~e.\
$\Theta$ and $\Theta + 2 n\pi$ are equivalent. In the following $\Theta = 0$
will be assumed.

The additional parameter $m$ introduces a bare mass for the gluino. This
mass in the fermionic sector breaks supersymmetry softly, i.~e.\ this kind
of breaking guarantees that the main features of the supersymmetric theory,
concerning the ultraviolet renormalisability, remain intact.

At zero temperature, gluons and gluinos can be found only in colourless
bound states. Those bound states are expected to form supermultiplets of
equal masses if exact supersymmetry is realised. A low-energy effective
Lagrangian has been formulated \cite{Veneziano:1982ah, Farrar:1997fn},
predicting a bound spectrum of mesons, glueballs and gluino-glueballs, which
has been subject of many numerical lattice investigations
\cite{Bergner:2012rv,Bergner:2013nwa}.

\section{Lattice discretisation}

On the lattice, the gauge fields $A_\mu^b(x)$ are associated with the links
of the lattice using the exponential map
\begin{equation}\label{expmap}
 U_\mu^R(x) = \exp{(\I g a A_\mu^b(x) \tau_b^R)},
\end{equation}
where $\tau_b^R$ are the Lie group generators in the representation $R$. In
the following, $U_\mu(x)$ and $V_\mu(x)$ will denote the link variables in
the fundamental and in the adjoint representation, respectively. The adjoint
links $V_\mu(x)$ are related to the fundamental links $U_\mu(x)$ through the
well-known formula
\begin{equation}
 V_\mu(x)_{ab} = 2\, \tr{(U_\mu(x)^\dag \tau_a^F U_\mu(x) \tau_b^F)}.
\end{equation}
In our investigations the gauge group is SU(2), therefore $U_\mu(x) \in$
SU(2) and $V_\mu(x) \in$ SO(3). The generators in the fundamental
representation are normalised such that:
\begin{equation}
 \tr (\tau_a^F \tau_b^F) = \frac{1}{2} \delta_{ab}.
\end{equation}

In our simulations, the gauge part of $S$ in \Eqref{continuum_action} is
discretised with a tree-level Symanzik improved action:
\begin{equation}
 S_g = \sum_x \re\, \tr \left\{\frac{\beta}{N_c}\sum_{\mu \neq \nu}
\left(\frac{5}{3}  P_{\mu\nu}(x) - \frac{1}{12} R_{\mu\nu}(x)\right)\right\},
\end{equation}
where $P_{\mu\nu}(x)$ is the standard plaquette term formed out of four
links, and $R_{\mu\nu}(x)$ represents a rectangle with lower left corner on
the point $x$. The gluino part of $S$ in \Eqref{continuum_action} is
represented on the lattice using the discretised version of the Dirac
operator, depending on the links in the adjoint representation $V_\mu(x)$:
\begin{equation}
 S_f = \sum_{x,y} \bar{\lambda}(y) D_W[V_\mu](y,x) \lambda(x).
\end{equation}
The action of the Wilson-Dirac operator $D_W$ on the gluino field $\lambda$
is given by (Dirac and colour indices suppressed)
\begin{equation}
 D_W(x,y)\lambda(y) = \lambda(x) - \kappa \sum_{\mu} 
\left\{ (1-\gamma_\mu) V_\mu(x) \lambda(x+\mu) 
+ (1+\gamma_\mu) V_\mu(x-\mu)^\dag \lambda(x-\mu)\right\},
\end{equation}
where $\kappa = \frac{1}{2m + 8}$ is the hopping parameter. Supersymmetry
and chiral symmetry are explicitly broken using this discretisation scheme.

Euclidean invariance is explicitly broken on the lattice and therefore it is
impossible to construct a local action invariant under supersymmetry
transformations for finite lattice spacing $a$
\cite{Bergner:2009vg,Kato:2008sp}. A fine tuning is needed to recover the
broken supersymmetry and chiral symmetry in the continuum limit. In
supersymmetric Yang-Mills theory the tuning of a single parameter, namely
the bare gluino mass $m$, is enough to recover both symmetries
\cite{Curci:1986sm,Suzuki:2012pc}. The tuning to the chiral limit can be
defined by the vanishing of the adjoint pion mass, which is defined in a
partially quenched setup \cite{Munster:2014cja}. We use it here to define
different lines of constant physics for theories with a softly broken
supersymmetry.

\section{The finite temperature phase diagram}

The $\mathcal{N}=1$ SYM is an asymptotically free theory, expected to behave
at high temperatures as a conformal gas of free gluons and gluinos
\cite{Amati:1988ft}. At zero temperature confinement and gluino condensation
take place. The possible phases are characterised by the expectation value
of their related order parameters considered as a function of the
temperature.

\subsection{Deconfinement phase transition}

A useful order parameter for the deconfinement transition is the Polyakov
loop
\begin{equation}
P_L = \frac{1}{V} \sum_{\vec{x}} 
\textrm{Tr}\left\{ \prod_{t=0}^{\Nt} U_4(\vec{x},t)\right\}.
\end{equation}
The expectation value of the Polyakov loop has the physical meaning of the
exponential of the negative free energy of a single static Dirac quark in
the fundamental representation
\begin{equation}
\langle P_L \rangle = \exp \left( - \frac{F_q}{T} \right).
\end{equation}
Therefore a non-vanishing value of $\langle P_L\rangle$ means that a state
with a single isolated quark exists, i.~e.\ deconfinement. Deconfinement is
associated with the spontaneous breaking of the center symmetry, defined by
the following transformation of the gauge fields in a fixed time-slice at
$t=t'$:
\begin{equation}
 U_4(\vec{x},t') \rightarrow U_4(\vec{x},t')' 
= \exp{\left(2\pi \I \frac{n}{N_c} \right)} U_4(\vec{x},t'),
\qquad n \in \{0, 1, \dots, N_c -1\}.
\end{equation}
In contrast to QCD with fermions in the fundamental representation, this
transformation leaves invariant both the gauge and the fermionic part of the
action of $\mathcal{N} = 1$ SYM. The Wilson-Dirac operator is written in
terms of links in the adjoint representation that are unaffected by the
complex rotation. On the other hand, the Polyakov loop transforms
non-trivially under the center transformations:
\begin{equation}
P_L \rightarrow P_L' = \exp{\left(2\pi \I \frac{n}{N_c} \right)} P_L\,.
\end{equation}
It is thus an exact order parameter for the deconfinement transition at any
value of the gluino mass $m$. The pattern for the center symmetry breaking
is hence the same as in pure SU($N_c$) gauge theories, and it is possible
that the Svetitsky-Yaffe conjecture \cite{Svetitsky:1982gs} is valid for
$\mathcal{N} = 1$ SYM. This conjecture implies a deconfinement transition of
second order for the gauge group SU(2), corresponding to the universality
class of the $Z_2$ Ising model in three dimensions.

\subsection{Chiral phase transition}

\begin{figure}[t]
\centering
\includegraphics[width=0.53\textwidth]{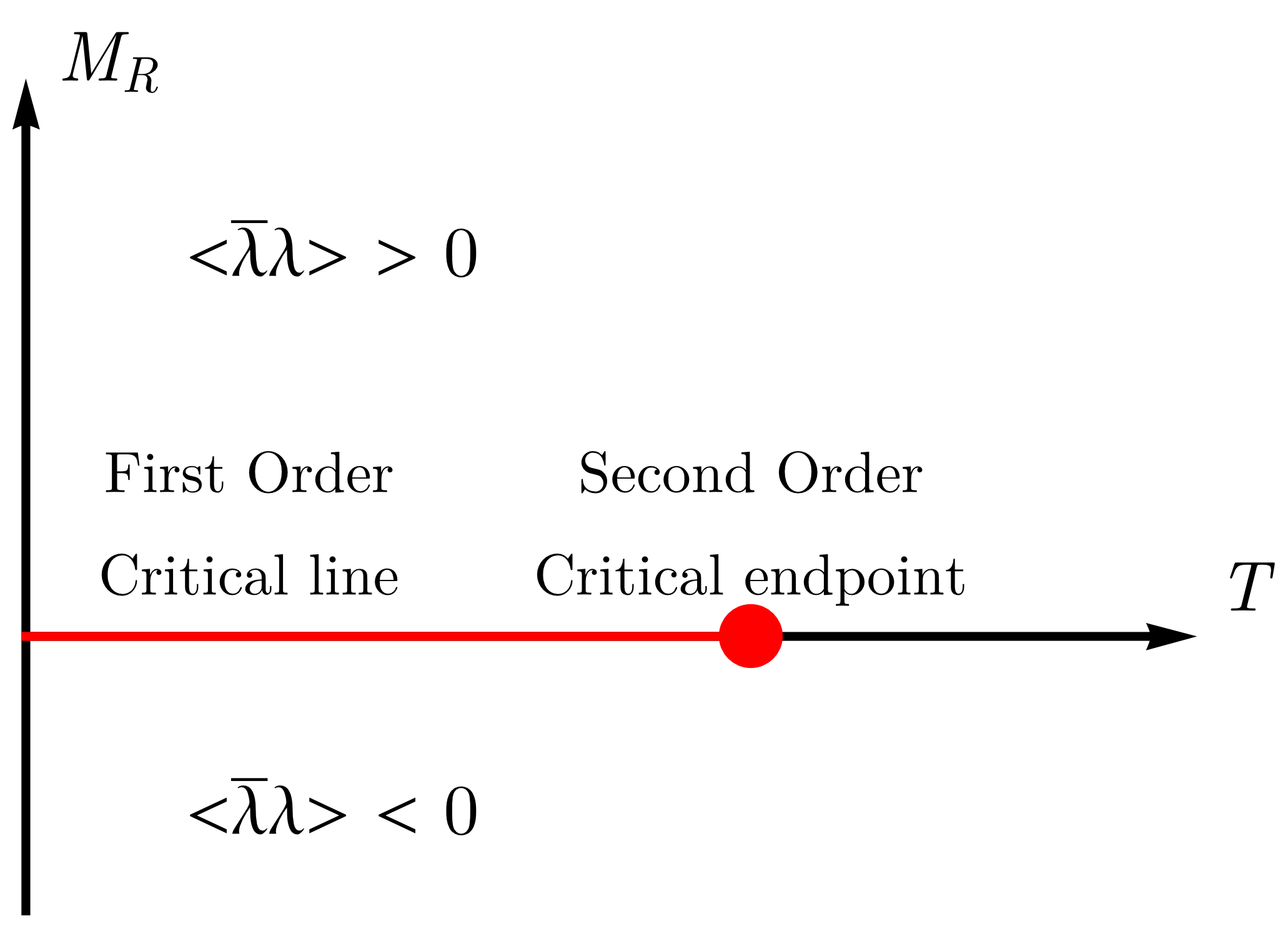}
\caption{Expected phase diagram for chiral symmetry breaking of
$\mathcal{N}=1$ SYM with gauge group SU(2). The chiral condensate $\langle
\bar{\lambda}\lambda \rangle$ is the order parameter of the chiral phase
transition. Each non-zero value of the renormalised gluino mass $M_R$
introduces a source of chiral symmetry breaking. At low temperatures, moving
from positive to negative gluino mass $M_R$, the chiral condensate jumps
from a positive to a negative expectation value. The phase transition is
therefore of first order with a coexistence of two phases. The line of first
order phase transitions is expected to terminate in a second order endpoint.
\label{chiral_phase_diagram}}
\end{figure}

The $\mathcal{N}=1$ SYM has a classical U(1)$_A$ axial symmetry, meaning
that the transformation
\begin{equation}
\lambda \rightarrow \lambda' = \exp{(- \I \omega \gamma_5)} \lambda
\end{equation}
leaves the action invariant when the gluino mass is exactly zero. This
symmetry is known as the R-symmetry U(1)$_R$ and it corresponds to the
relative rotation of the left- and the right-handed Weyl components of the
gluino field $\lambda$.

Beyond the classical level, quantum fluctuations break chiral symmetry by a
term proportional to the gauge coupling and to the number of colours:
\begin{equation}
\partial_\mu J^\mu_5 = \partial_\mu (\bar{\lambda} \gamma^\mu \gamma_5 \lambda) 
= N_c \frac{g^2}{32\pi^2} \epsilon_{\mu\nu\rho\sigma}F^{\mu\nu}F^{\rho\sigma}.
\end{equation}
The dependence on $N_c$ is absent in QCD, and it is typical of gauge models
with fermions in the adjoint representation.

The anomalous contribution to the axial transformations can be absorbed in
the periodicity of the parameter $\Theta$:
\begin{equation}
 \Theta \rightarrow \Theta - 2 N_c \,\omega,
\end{equation}
if the angle $\omega$ assumes one of the values $\omega = \frac{n
\pi}{N_c}$, $n=0, \ldots, 2N_c - 1$. The remaining chiral symmetries thus
from the group $Z_{2N_c}$. Numerical investigations \cite{Kirchner:1998mp}
have confirmed the conjecture
\cite{Amati:1988ft,Seiberg:1994bz,Seiberg:1994pq} that this invariance is
spontaneously broken at zero temperature by a non-vanishing expectation
value of the gluino condensate $\langle\bar{\lambda}\lambda\rangle \neq 0$
to a remaining $Z_2$ symmetry corresponding to the sign flip $\lambda
\rightarrow - \lambda$. The complete pattern of chiral symmetry breaking is
thus
\begin{equation}
\mathrm{U}(1)_A \rightarrow Z_{2N_c} \rightarrow Z_2 .
\end{equation}

\begin{figure}[t]
\centering
\subfigure[Coincident phase transitions]{
\includegraphics[width=0.47\textwidth]{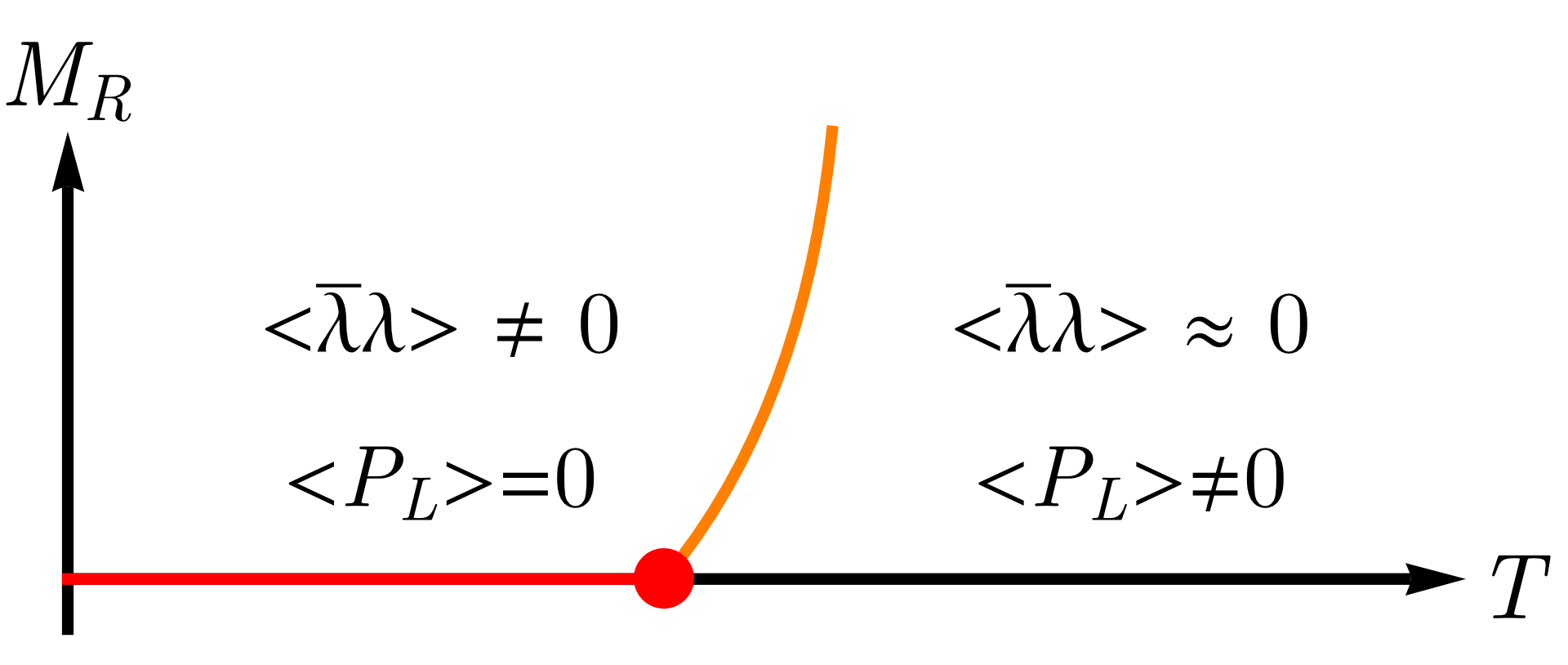}}
\subfigure[Mixed phases allowed]{
\includegraphics[width=0.47\textwidth]{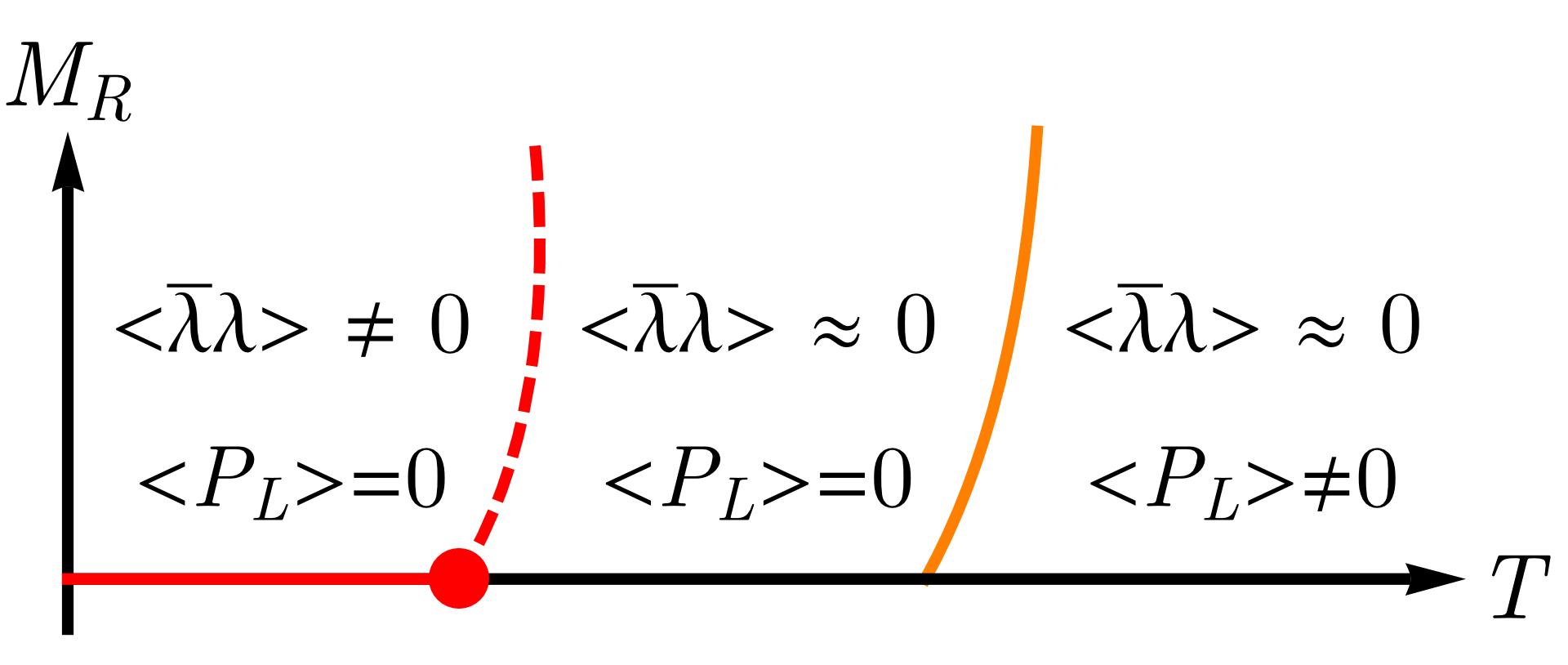}}
\subfigure[Mixed phases allowed]{
\includegraphics[width=0.47\textwidth]{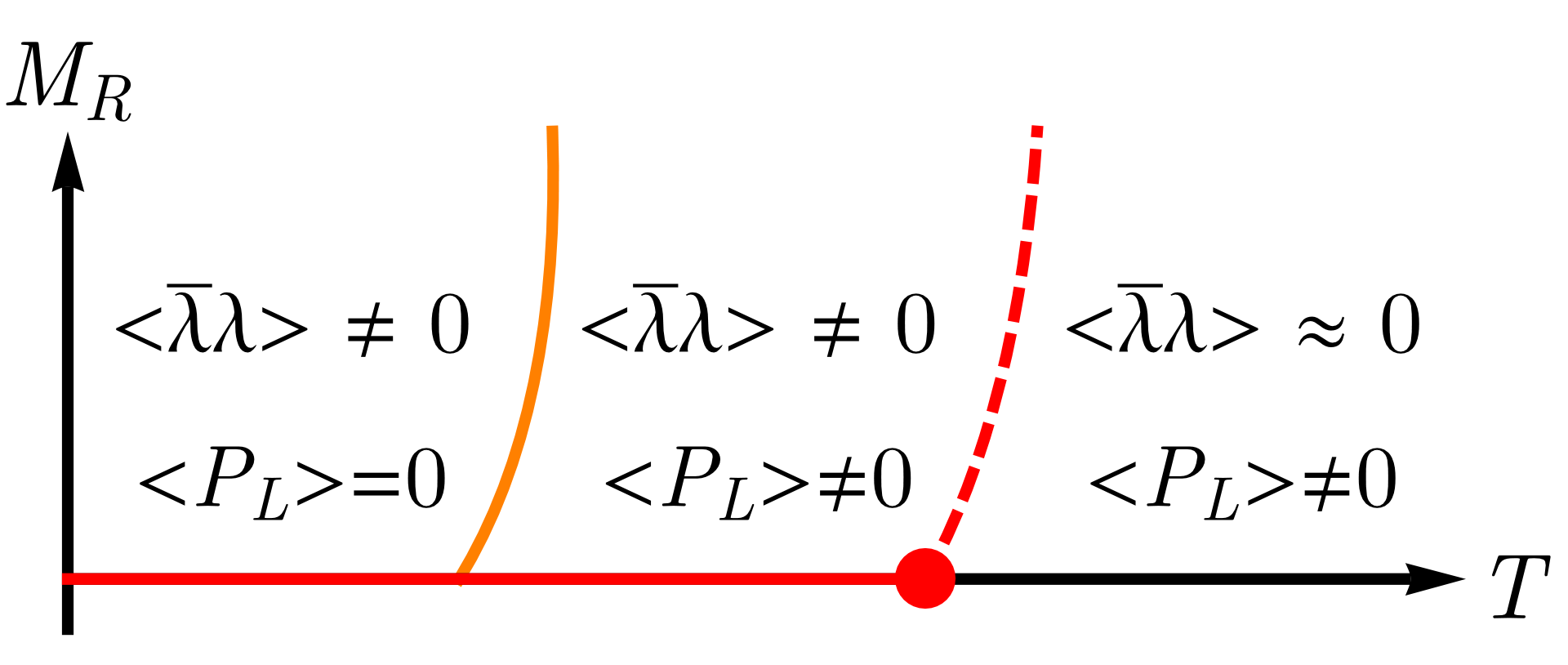}}
\caption{Possible scenarios for the phase diagram of $\mathcal{N}=1$ SYM.
The two chiral phases are separated by a crossover for $M_R \neq 0$ (dashed
red lines) and by a phase transition in the massless limit. The orange line
represents the deconfinement phase transition, present for any value of
$M_R$. (a) In the supersymmetric limit chiral and deconfinement transition
coincide. (b) A mixed confined phase occurs with chiral symmetry restored.
(c) A mixed deconfined phase occurs with chiral symmetry broken.
\label{scenarios}}
\end{figure}

The phase transition associated with the spontaneous breaking $Z_{2N_c}
\rightarrow Z_2$ is of first order at zero temperature, related to the jump
of the expectation value of the chiral condensate. The system is in this
respect similar to a $Z_{N_c}$ Ising model ($Z_{N_c}=Z_{2 N_c}/Z_{2}$), with
the gluino condensate corresponding to the spontaneous magnetisation and the
renormalised gluino mass to the external magnetic field. This similarity
suggests that for the gauge group SU(2) there is a critical temperature
$T_{c}^{\textrm{chiral}}$ of a second order phase transition and a phase
with restored $Z_4$ symmetry at high temperatures, see
Fig.~\ref{chiral_phase_diagram}.

There are three possible scenarios for the relation of deconfinement and
chiral symmetry restoration. $T_{c}^{\textrm{chiral}}$ might coincide with
the deconfinement transition temperature $T_{c}^{\textrm{deconf.}}$, but
there is no restriction to this scenario from first principles. If they do
not coincide either a mixed deconfined phase with broken chiral symmetry or
a mixed confined phase with restored chiral symmetry exists, see
Fig.~\ref{scenarios}.

In addition, the remaining part of the U(1)$_R$ symmetry broken by the
anomaly could be effectively restored in high temperature limit.

\section{Simulation algorithms}

In order to perform Monte Carlo simulations of  $\mathcal{N}=1$ SYM, the
gluino field is integrated out in the path integral. For Majorana fermions
the result is the Pfaffian of the Wilson-Dirac operator
\begin{equation}
 Z = \int D U\ \pf ( C D_W ) \exp{(-S_g)}.
\end{equation}
The Pfaffian of an antisymmetric matrix is related to the square root of the
determinant by
\begin{equation}
 \pf ( C D_W ) = \sign(\pf ( C D_W )) \sqrt{\dt(D_W)}.
\end{equation}
The additional factor leads to the notorious sign problem of this theory. At
a fixed lattice spacing, configurations with a negative Pfaffian sign can
appear. This happens in particular at small residual gluino masses close to
the supersymmetric limit. These contributions are reduced moving to smaller
lattice spacings and the Pfaffian is strictly positive in the continuum
limit. It is hence possible to stay in the region where the sign problem is
irrelevant. On the other hand, a reliable extrapolation to the
supersymmetric limit requires small gluino masses, and negative Pfaffian
signs cannot be excluded. We have monitored the Pfaffian signs for the runs
with the most critical parameters using the method introduced in
\cite{Bergner:2011zp} to keep this effect under control.

Our simulations have been performed using the Hybrid Monte Carlo algorithm
(HMC). We have applied two different approaches for the approximation of the
square root of the determinant: an exploratory study was done with a code
based on the polynomial (PHMC) approximation; the second, and main part of
the work, was performed instead using a new code, based on the rational
(RHMC) approximation. The PHMC algorithm is used with one-level stout links
in the Wilson-Dirac operator, while the RHMC is used with unsmeared links.

When the renormalised gluino mass is sent to zero, the Wilson-Dirac operator
becomes ill-conditioned and the computational demand for the numerical
integration of the classical trajectory in the HMC increases drastically.
The most reliable approach is therefore to perform the simulations for
several non-zero values of the gluino mass and obtain the supersymmetric
limit by extrapolation of the results.

\section{Scale setting in supersymmetric Yang-Mills theory}
\label{sec:scale}

The phase diagram of $\mathcal{N}=1$ SYM theory is investigated on lattices
of finite size $N_s^3\times \Nt$ for different values of the bare couplings
$\kappa$ and $\beta$. The boundary conditions are anti-periodic in the
Euclidean time direction for fermions and periodic in all other cases. In
order to convert the bare parameters into physical units the size of the
lattice spacing in physical units is needed. This is done by means of the
Sommer parameter $r_0$ \cite{Sommer:1993ce} and the $w_0$
parameter~\cite{Borsanyi:2012zs}. The calculation of the scale is based on
results of simulations at zero temperature.

At fixed $\beta$ and $\kappa$ the temperature is proportional to the inverse
of the number of lattice sites in the temporal direction $\Nt$,
\begin{equation}
 T = \frac{1}{\Nt a}.
\end{equation}
The continuum limit is obtained when $a \rightarrow 0$ at fixed $T$. The
phase transitions are determined as a function of the renormalised value of
the residual gluino mass $M_R$ that breaks supersymmetry softly. The
extrapolation of the transition temperatures to the supersymmetric limit
$M_R\rightarrow 0$ is the final result of our calculation. It has been shown
in a partially quenched setup that $M_R$ is proportional to the square of
the adjoint pion mass $m_{\api}$ \cite{Munster:2014cja}. For several
critical couplings $\bcd$ and $\kcd$ of the deconfinement transition the
adjoint pion mass is measured in zero temperature simulations on lattices of
size $N_s^3\times 2 N_s$. These results are summarised in
Table~\ref{tablemeaszero}. At small values of $m_{\api}$ the dependence of
$M_R$ on $1/\kappa$ for fixed $\beta$ is approximately linear. In cases,
where zero temperature results at $\bcd$ and different values of $\kappa$
were present from previous investigations we apply a linear fit to
interpolate $M_R$ at $\kcd$.

In the continuum the scaling function is given by the
Novikov-Shifman-Vainshtein-Zakharov beta-function \cite{Novikov:1983uc} to
all orders in perturbation theory. From this beta-function the one-loop
perturbative scaling of the lattice spacing as a function of the coupling
$\beta$ in the supersymmetric limit is given by
\begin{equation}
 a(\beta) = \frac{1}{\Lambda} \exp{\left(-\frac{\pi^2}{3} \beta \right)}.
\end{equation}
However, at finite lattice spacings it is more feasible to consider a
non-perturbative scale setting based on measurable scale parameters.

We consider two different observables for the scale setting. The first one
is the Sommer parameter $r_0/a$ obtained from the static quark-antiquark
potential \cite{Sommer:1993ce}. Since in SYM there is no string breaking for
static quarks in the fundamental representation, this scale can be measured
in the same way as for pure Yang-Mills theory. The second observable is the
recently proposed alternative $w_0/a$ \cite{Borsanyi:2012zs}. This
observable is obtained from the gradient flow of the gauge action density
$E$ with $G^a_{\mu\nu}$ represented by clover plaquettes,
\begin{equation}
 E=\frac{1}{4}G_{\mu\nu}^aG_{\mu\nu}^a\; .
\end{equation}
The gradient flow is defined as continuous smearing procedure using, in our
case, the functional derivative of the Wilson plaquette action. The scale
parameter $w_0/a$ is defined by the flow time $t$, where
\begin{equation}
 \left. t\frac{d}{dt}\left[ t^2 \langle E(t) \rangle \right]\right|_{t=w_0^2}=0.3\; .
\end{equation}

\begin{figure}[tb]
\centering
\subfigure[Wilson flow energy]{%
\includegraphics[width=0.45\textwidth]{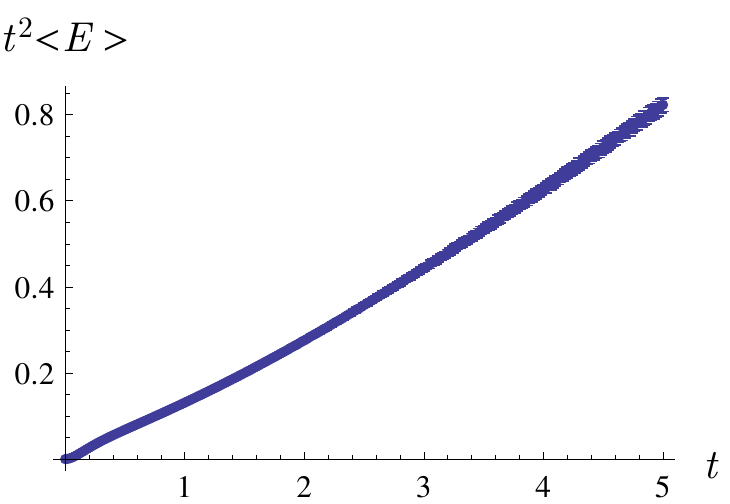}
\label{wilson_flow_integrated}}
\subfigure[Mass dependence of $w_0/a$]{%
\includegraphics[width=0.51\textwidth]{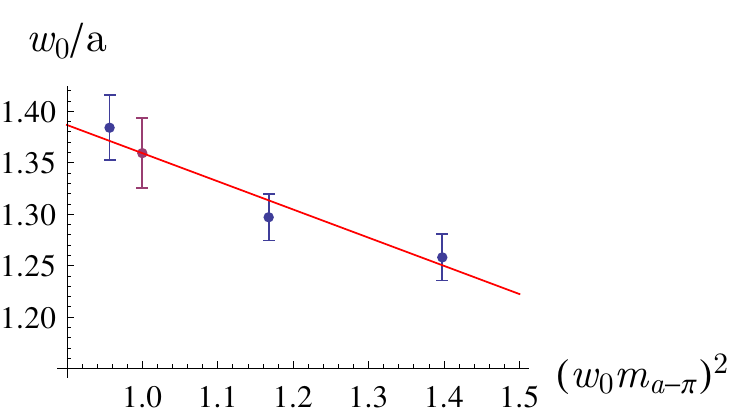}
\label{w0_mass_dependence}}
\caption{a) Expectation value of the gauge energy as a function of the
Wilson flow time $t$ on a $14^4$ lattice at $\beta=1.65$ and $\kappa =
0.1875$. b) Dependence of $w_0/a$ on the gluino mass for $\beta = 1.62$, see
Table~\ref{tablemeaszero}. In a \emph{mass independent} renormalisation
scheme, the scale is fixed at a reference gluino mass (in our case $(w_0
m_{\api})^2 = 1$, purple point), extracted from a linear fit of the
available data.}
\end{figure}

The dependence of the observable on the flow time is shown in
Fig.~\ref{wilson_flow_integrated}. As expected, the dependence of $t^2 E$ on
$t$ is approximately linear for large $t$.

An advantage of $r_0/a$ and $w_0/a$ is their weak dependence on the residual
gluino mass. In both cases a mild, but not negligible, linear dependence is
observed, see Fig.~\ref{w0_mass_dependence}. As a consequence two different
approaches can be applied to set the scale: in a \emph{mass dependent}
renormalisation scheme the scale is set separately at each value of $M_R$
and consequently the lattice spacing depends on the gluino mass,
\begin{equation}
 a \equiv a(\beta,M_R).
\end{equation}
In the second approach the lattice spacing is taken to be independent of the
gluino mass $M_R$,
\begin{equation}
 a \equiv a(\beta)\; .
\end{equation}
Therefore, in this case the linear behaviour of $w_0/a$ and $r_0/a$ is
interpreted as a physical dependence of their value on the fermion mass
\cite{De:2008xt}. This approach is called \emph{mass independent}
renormalisation scheme and it requires an extrapolation of $r_0/a$ and
$w_0/a$ to a fixed reference value of $M_R$, as shown in
Fig.~\ref{w0_mass_dependence}.

\begin{figure}
\centering
\includegraphics[width=0.7\textwidth]{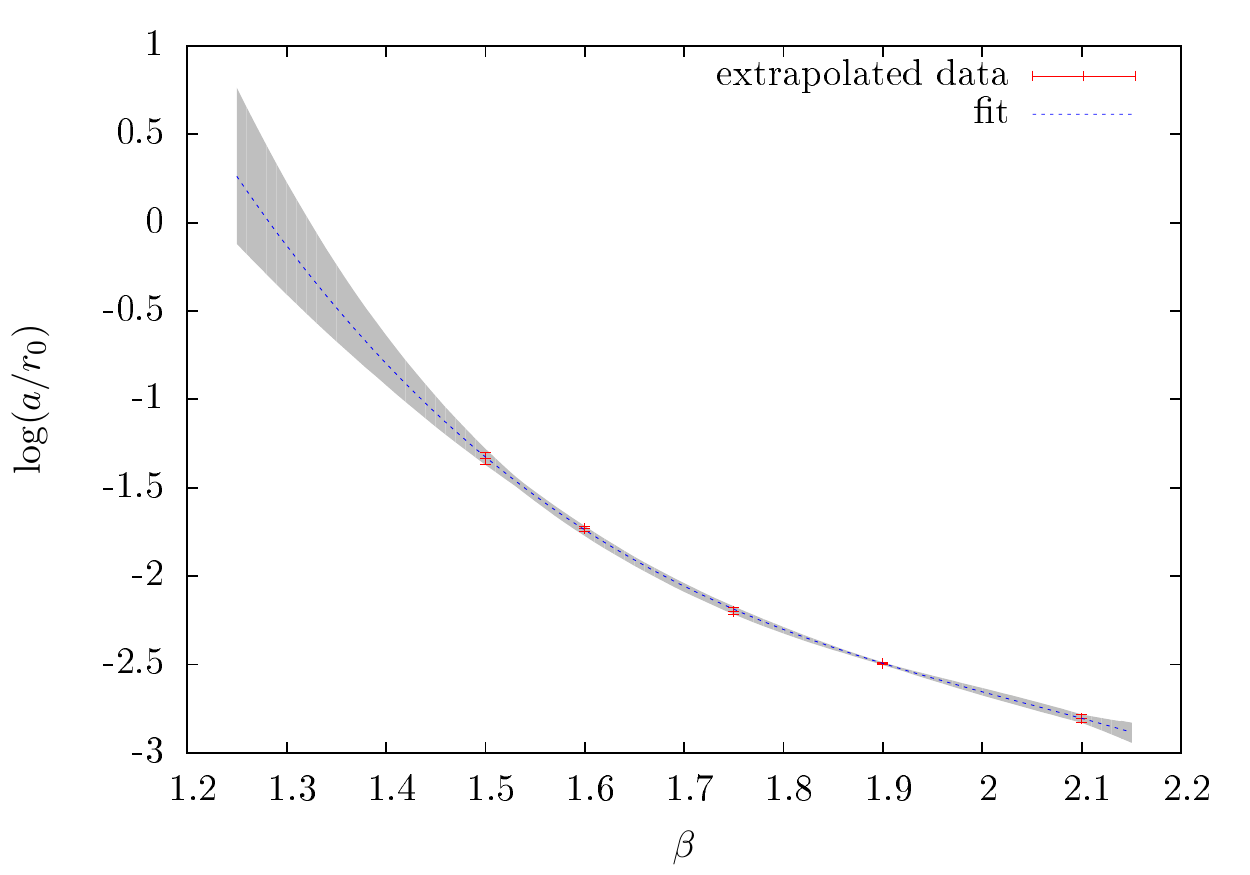}
\caption{The scale setting of supersymmetric Yang-Mills theory. The figure
shows the values of $r_0/a$ extrapolated to the chiral limit ($ M_R=0$).
\label{fig:scalesetting1lstout}}
\end{figure}

In our exploratory study, where the PHMC approximation with one-level of
stout smearing has been used, the mass independent approach has been
applied. {}From previous investigations\footnote{The details of the
simulations performed at $\beta=1.9$ and $\beta=2.1$ will be presented in a
paper which is in preparation.\label{note1}} we have already some zero
temperature results at $\beta=1.6$, $1.75$, $1.9$, and $2.1$, see
Table~\ref{betar0values}. In these studies the Sommer parameter $r_0/a$ has
been extrapolated to the value $\rne/a$ at the supersymmetric limit,
corresponding to a reference scale of $M_R=0$. We have completed the data
with additional simulations at $\beta=1.5$, see Table~\ref{tablemeasstout2}.
The dependence of the scale $\rne/a$ on $\beta$ is fitted with a similar
parametrisation as used in \cite{Necco:2001xg},
\begin{equation}
 \log(a/\rne)=a_1+a_2(\beta-2)+a_3(\beta-2)^2+a_4(\beta-2)^3\; .
\label{interpolatingeq}
\end{equation}
Due to the limited amount of data the error of the coefficients $a_1$,
$a_2$, $a_3$, and $a_4$ is not reliably obtained from a single fit. To get a
better estimate we have assembled several samples of data points by taking
values within the given error bounds of each point. With the fits of these
samples one obtains a set of curves that determines the error bound of the
interpolation, see Fig.~\ref{fig:scalesetting1lstout}. This result
determines the values of $\rne/a$ for the $\beta$ values without enough data
from zero temperature simulations.

In the second part of the work, where the simulations have been performed
with the RHMC algorithm and without stout smearing, the scale has been set
by the parameter $w_0$ and both the mass dependent and independent schemes.
In the mass independent scheme $w_0$ is extrapolated to the chosen reference
point $(w_0 m_{\api})^2 = 1$, see Fig.~\ref{w0_mass_dependence}. This
reference point can be accurately extrapolated already from a small number
of zero temperature simulations. The obtained value $w_0^e(\beta)$ is used
to fix the scale for all the simulations with the same value of $\beta$. In
order to exclude possible systematic errors with this approach we have also
applied a mass dependent scale setting prescription, see
Table~\ref{tablemeaszero}. In that case the scale at $\beta$ and $\kappa$ is
set with the value of $w_0$ measured at the same combination of the
parameters in a zero temperature simulation.

\section{The confinement-deconfinement phase transition}

\begin{figure}
\centering
\subfigure[Temperature versus $\kappa$]
{\includegraphics[width=0.7\textwidth]{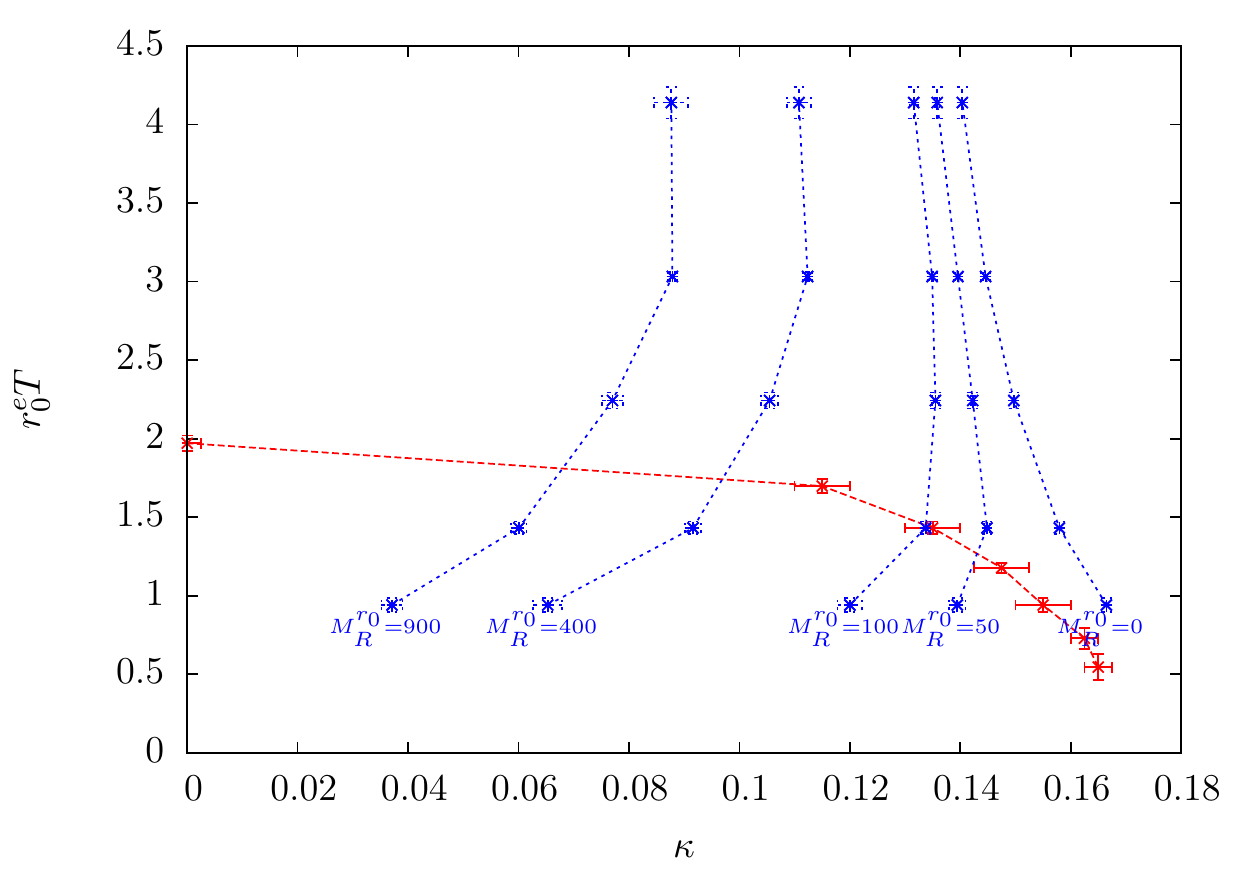}
\label{fig:roughps-a}}
\subfigure[Extrapolation of $T_c$]
{\includegraphics[width=0.7\textwidth]{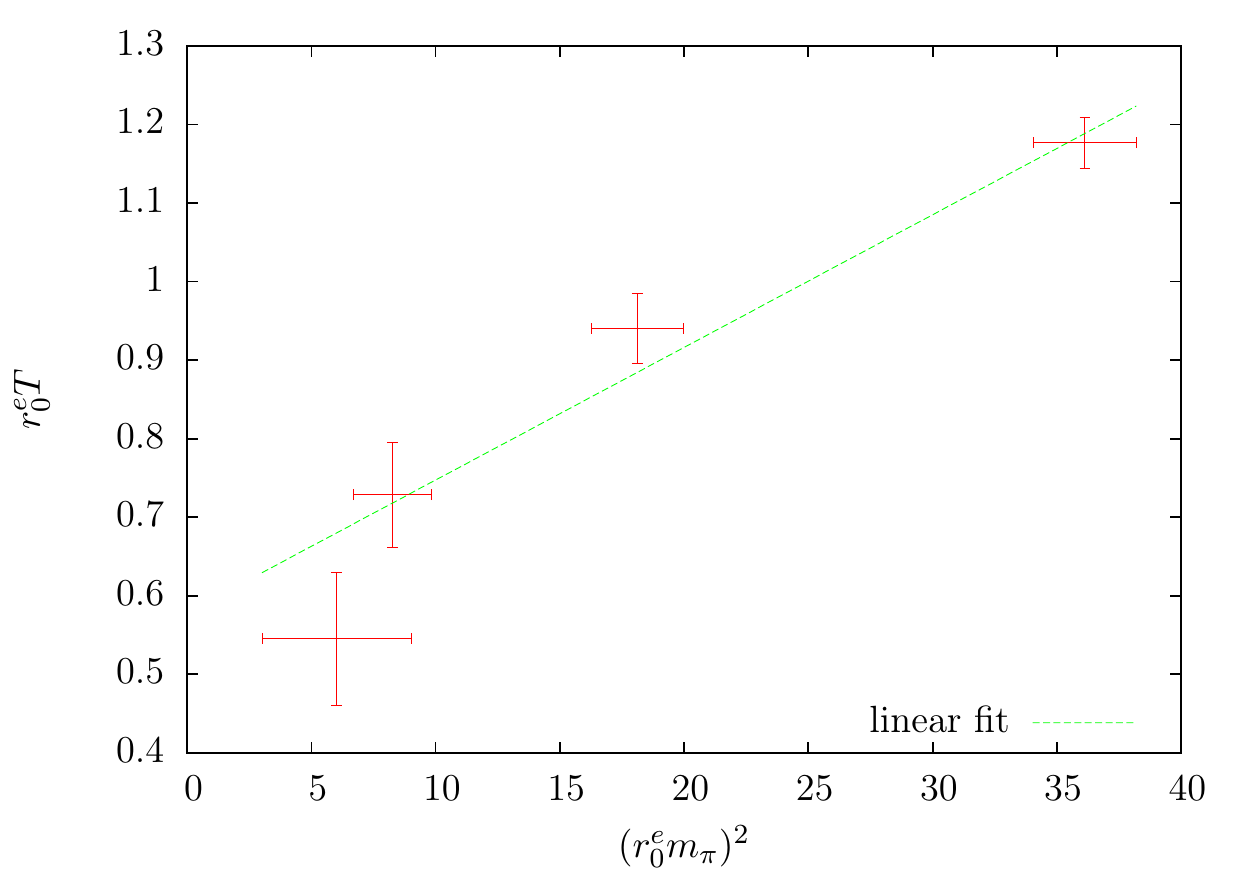}
\label{fig:roughps-b}}
\caption{a) The confinement-deconfinement phase transition as a function
of the bare parameter $\kappa$ (red line). The simulations are done with one level of
stout smearing on $8^3\times 4$ and $12^3\times 4$ lattices. The blue lines
indicate the lines of constant physics, corresponding to a fixed residual
mass $M^{r_0}_R=(\rne m_{\api})^2$. The chiral limit is approached at the
line $M_R=0$. In these data the scale $\rne$ is fixed to the extrapolated
value in the chiral limit. b) Linear extrapolation of the critical 
temperature to the chiral limit.}
\end{figure}
\begin{figure}[tb]
\centering
\subfigure[Polyakov loop $\langle P_L \rangle$]
{\includegraphics[width=0.48\textwidth]{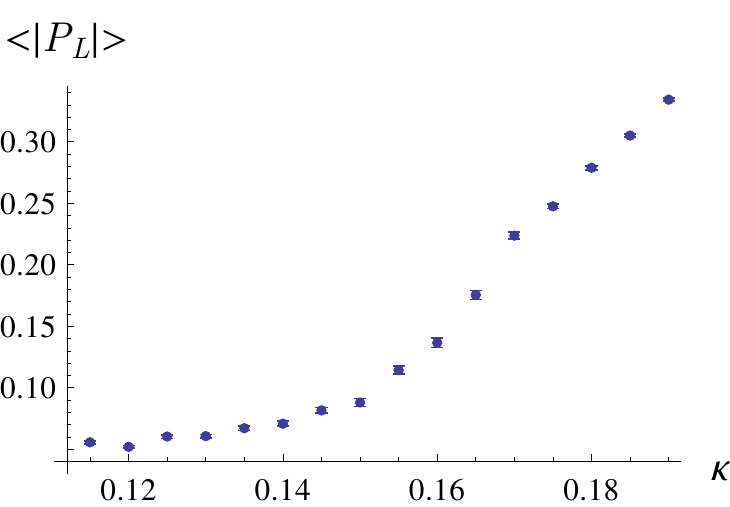}
\label{polyakov4_12}}
\subfigure[Polyakov loop susceptibility $\chi_P$]
{\includegraphics[width=0.48\textwidth]{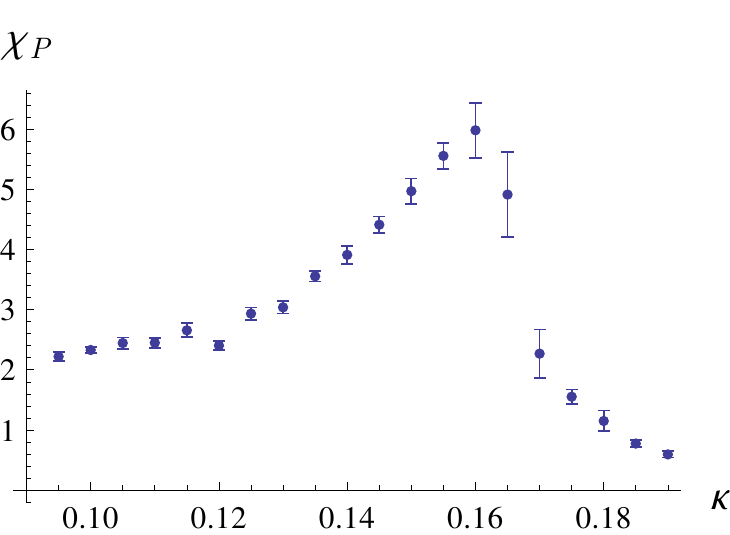}
\label{susc4_12}}
\caption{The expectation value of the Polyakov loop and of its
susceptibility on a $12^3\times4$ lattice at $\beta =1.65$.}
\end{figure}
We have done the first scan of the phase diagram of supersymmetric
Yang-Mills theory with the same parameters and settings as in our zero
temperature studies of the particle spectrum \cite{Bergner:2013nwa}. One
level of stout smearing has been applied in these simulations. They have
been performed at fixed lattice size of $N_s=8$ and $\Nt=4$. A few runs, to
check the finite volume effects, have been done using $N_s=12$. We have
collected for each value of $\beta$ the critical value $\kcd$ determined by
the peak in the Polyakov loop susceptibility. The results can be found in
Table~\ref{tablemeasstout} and are represented by the red symbols in
Fig.~\ref{fig:roughps-a}. The value of $\rne T=(\rne/a)/\Nt$ is calculated
with the value of $\rne/a$, for each value of $\beta$, obtained from an
interpolation based on \Eqref{interpolatingeq}.

The red line indicates the transition between confined and deconfined phase
for a residual gluino mass $M_R$ different from zero. In the limit of
$\kappa=0$, i.~e.\ infinite $M_R$, the result of pure SU(2) Yang-Mills
theory is obtained. Lowering the mass, i.e. increasing $\kappa$, towards the
chiral limit the phase transition temperature decreases. Since $\Nt$ is
fixed these lower temperatures correspond to a smaller value of $\beta$ and
a larger lattice spacing.

The blue symbols in Fig.~\ref{fig:roughps-a} indicate the lines of constant
$M^{r_0}_R\doteq (\rne m_{\api})^2$ in zero temperature simulations. They
are based on the results of simulations at $\beta=1.5$, $1.6$, $1.75$,
$1.9$, and $2.1$, where theses five values correspond to the five points
along each blue line. A linear interpolation of $m_{\api}$ as a function of
$1/\kappa$ has been used. Each of the intersections between the (blue) lines of
constant $M_R$ and the phase transition (red) line corresponds to the phase
transition at temperature $\rne T_c$ of a theory with softly broken
supersymmetry.

The phase transition of the supersymmetric Yang-Mills theory would
correspond to the intersection between the red line and the $M_R=0$ line.
{}From these results it can only be estimated to be around $ 0.5 \lesssim
\rne T_c \lesssim 1.0$. A systematic extrapolation can be done as a function
of the physical parameter $M_R$ instead of the bare parameter $\kappa$. We
take the four largest values of $\kappa$, corresponding to $\beta=1.55,
1.50, 1.45, 1.40$. They are converted to $M_R^{r_0}$ using the values of
$am_{\api}$ of Table~\ref{tablemeasstout2}. We can perform a linear fit to
determine the critical temperature, see Fig.~\ref{fig:roughps-b}. Already
with these rough data the phase transition point can thus be estimated to be
around
\begin{equation}\label{eq:roughstout1}
 \rne T_c= 0.577(81)\; .
\end{equation}

In these first investigations it turned out that much larger statistics and
a more precise scale estimation is necessary. Compared to these
uncertainties the improvement by stout smearing of the links is not
important. We have therefore developed a new more flexible update program
and performed a careful investigation in the region with small $M_R$ at the
phase transition. In that way we have obtained a more reliable extrapolation
of the supersymmetric limit without stout smearing.

As explained above, the parameter $w_0$ is used as a more recent alternative
scale setting. Different spatial and temporal lattice extents are considered
with lattice sizes $N_s^3 \times \Nt$ = $8^3\times 4$, $12^3\times 4$,
${16^3\times4}$, and $10^3 \times 5$ to estimate the influence of finite
size effects and lattice artifacts. For each lattice size, simulations are
done with different bare gauge couplings $\beta$ and gluino masses
($\kappa$). The details of the simulations, done at the critical value
$\kcd$, are summarised in Table~\ref{tablemeasunstout}. The autocorrelations
between consecutive configurations generated by the HMC algorithm increase
drastically near the critical point of the deconfinement transition. We have
increased the statistics near the phase transition in order to compensate
this effect, and we have investigated accurately the finite volume effects and
the scaling behaviour. Hence these points require a huge amount of computer
time.

The bare gluino mass ($\kappa$) is varied for each $\Nt$ and for each
$\beta$ to locate the point of the deconfinement phase transition $\kcd$.
Fig.~\ref{polyakov4_12} demonstrates this approach for a lattice size
$12^3\times 4$ and $\beta = 1.65$, where the Polyakov loop starts to rise at
$\kcd \simeq 0.15$. The point of the phase transition can be determined more
clearly by the maximum of the Polyakov loop susceptibility
\begin{equation}
\label{eq:sus}
 \chi_P = V (\langle |P_L| ^2 \rangle - \langle |P_L|  \rangle^2).
\end{equation}
Here the susceptibility is defined in terms of the modulus of the Polyakov
loop. While this choice does not alter the position of the peak, it
introduces a non-zero value of $\chi_{P}$ below the critical point.

As can be seen in Fig.~\ref{susc4_12}, the location of the transition is
found at $\kcd = 0.160(5)$, where the susceptibility shows a clear peak. We
have found that $\kcd$ defined in this way has only a mild finite volume
dependence, which is impossible to distinguish with our current precision.

\begin{figure}[tb]
\centering
\includegraphics[width=0.67\textwidth]{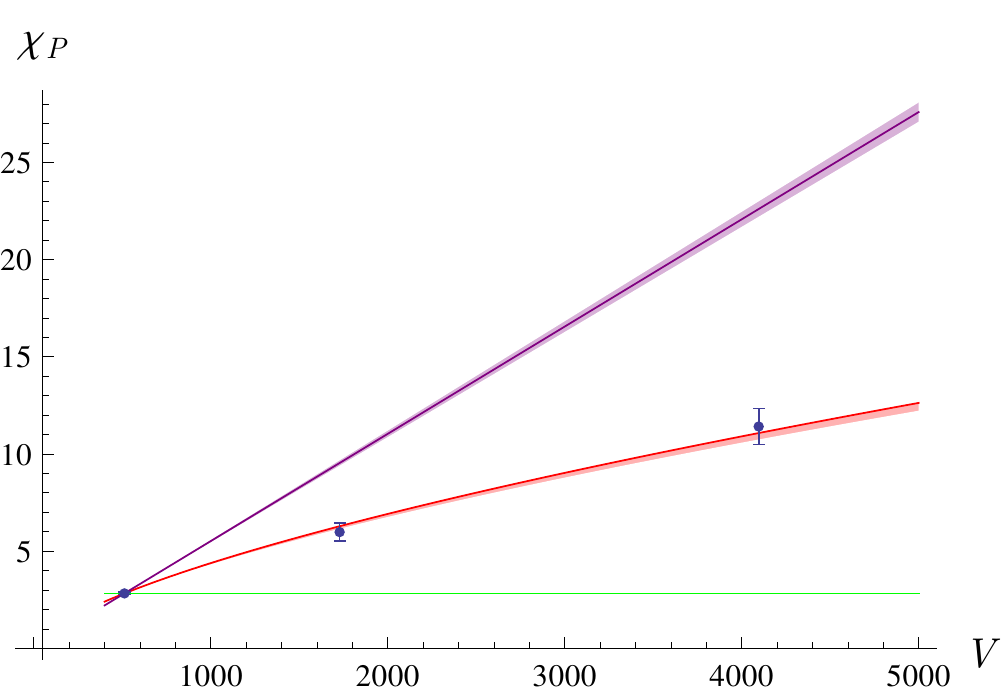}
\caption{Finite size scaling for the susceptibility $\chi(|P_L|)$ at
$\beta=1.65$ and $\kappa=0.160$. The red line is the expected scaling for a
second order phase transition in the universality class of the $Z_2$ Ising
model, the purple and the green lines represent instead a first order phase
transition and a cross-over, respectively. The coloured shadows indicate the
errors from the extrapolation of \Eqref{fssformula}.
\label{fss}}
\end{figure}

\begin{figure}[tb]
\centering
\subfigure[$\kappa=0.100$]{
\includegraphics[width=0.27\textwidth]{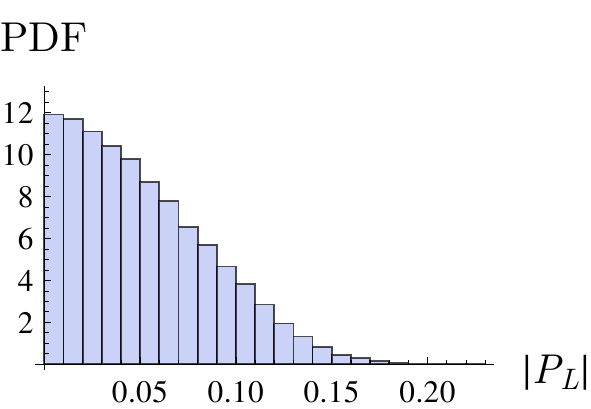}}
\subfigure[$\kappa=0.125$]{
\includegraphics[width=0.27\textwidth]{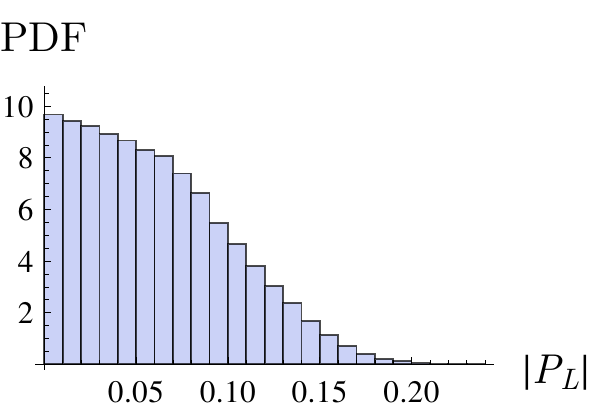}}
\subfigure[$\kappa=0.145$]{
\includegraphics[width=0.27\textwidth]{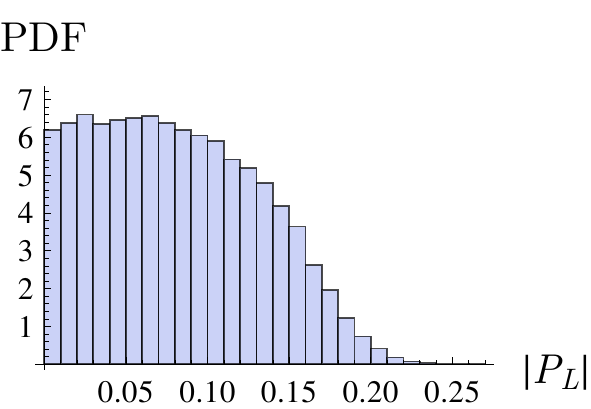}}
\subfigure[$\kappa=0.150$]{
\includegraphics[width=0.27\textwidth]{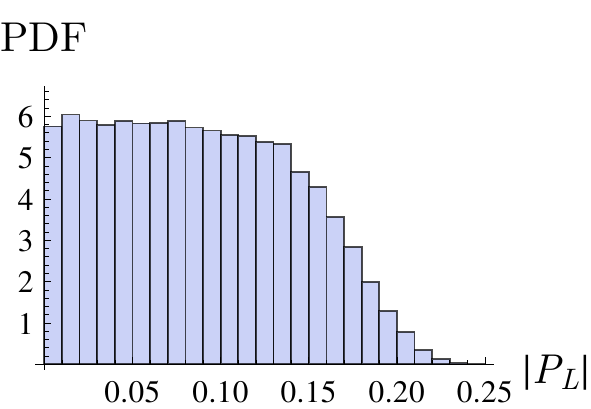}}
\subfigure[$\kappa=0.160$]{
\includegraphics[width=0.27\textwidth]{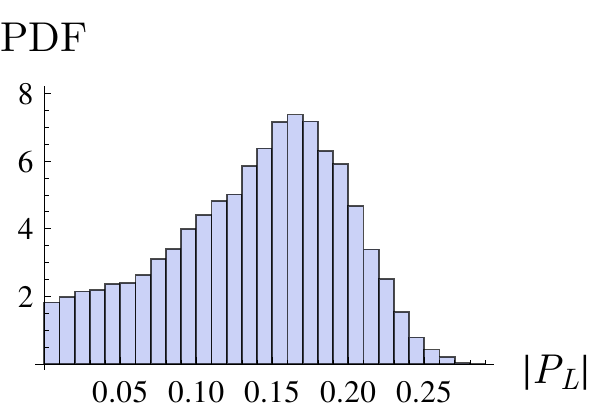}}
\subfigure[$\kappa=0.165$]{
\includegraphics[width=0.27\textwidth]{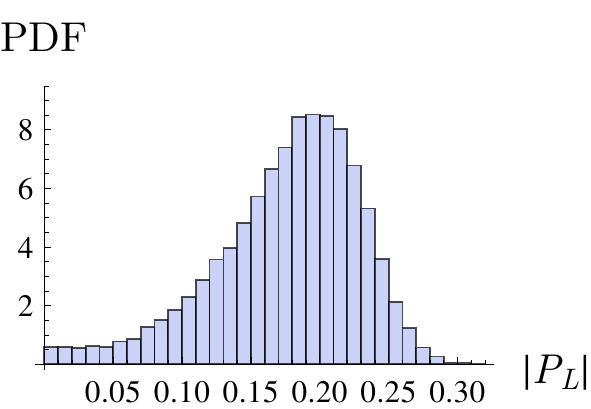}}
\caption{Polyakov loop distribution on a $12^3\times4$ lattice at $\beta =
1.65$ and various $\kappa$.
\label{distribution}}
\end{figure}

The finite size scaling of the susceptibility $\chi_{P}$ contains
information about the nature of the phase transition in the infinite volume
limit. The susceptibility has a scaling dependence on the volume,
\begin{equation}\label{fssformula}
 \frac{\chi_P(V_1)}{\chi_P(V_2)} = \left(\frac{V_1}{V_2}\right)^x ,
\end{equation}
which is linear for a first order phase transition ($x= 1$), flat for a
crossover behaviour ($x=0$), and non-linear, with $x = 0.657(4)$, for a
second order phase transition in the universality class of three-dimensional
$Z_2$ Ising model~\cite{Ferrenberg1991}. We have obtained the ratios of the
Polyakov loop susceptibility for $V_2 = 8^3$ and $V_1 = 12^3, 16^3$ with
$\Nt=4$. The resulting volume dependence is shown in Fig.~\ref{fss}. The
Svetitsky-Yaffe conjecture is in good agreement with the data and a possible
change from the second order phase transition of pure gauge theory to first
order induced by gluinos seems to be excluded.

As a further evidence for this statement the distributions of the Polyakov
loop at different values of $\kappa$ demonstrate the slow continuous
emergence of a new peak in addition to the central distribution of the
absolute value, see Fig.~\ref{distribution}. This is in accordance with the
divergence of the correlation length at a second order phase transition.

\begin{figure}[t]
\centering
\subfigure[Mass independent]
{\includegraphics[width=0.48\textwidth]{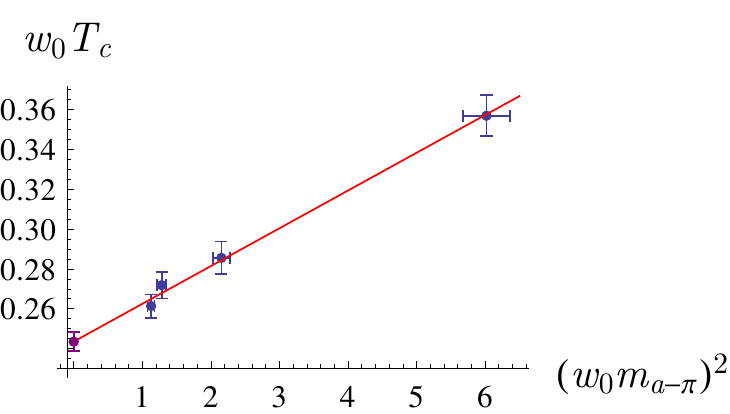}
\label{extrapolation-mass-independent}}
\subfigure[Mass dependent]
{\includegraphics[width=0.48\textwidth]{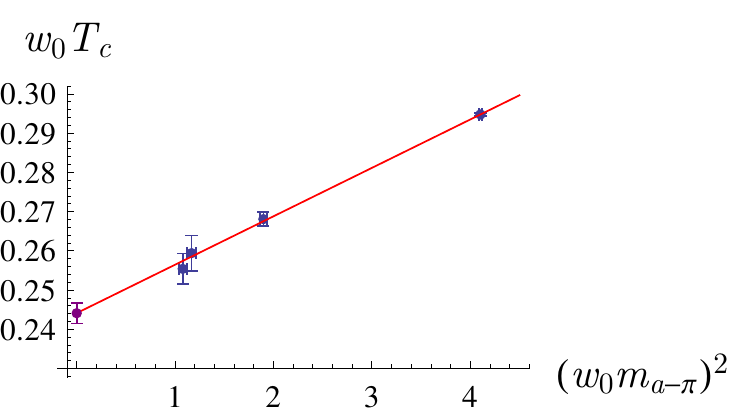}
\label{extrapolation-mass-dependent}}
\caption{The critical temperature of the deconfinement phase transition is
extrapolated to the supersymmetric limit by extrapolating the results to the
point where $(w_0 m_{\api})^2$ is equal to zero. The four points plotted in
the two figures can be found in Table~\ref{tablemeaszero}.}
\end{figure}

The peak of the Polyakov loop susceptibility, \Eqref{eq:sus}, defines a
critical combination of bare parameters $(\kcd , \bcd)$, see
Table~\ref{tablemeasunstout}. At these values new simulations have been
performed at zero temperature, see Table~\ref{tablemeaszero}, to determine
the value of the adjoint pion mass in lattice units $a m_{\api}$ and to set
the scale. Note that in Table~\ref{tablemeaszero}, for each $\bcd$, in
addition to the result obtained at the corresponding $\kcd$, other values
determined at different $\kappa$ are present: these are necessary to
extrapolate the mass independent scale $w_0^e/a$ as explained in
Sec.~\ref{sec:scale}. With the previous determinations, the critical
temperature and the adjoint pion mass are obtained in dimensionless units,
i.~e.\ $w_0^e T_c$ and $w_0^e m_{\api}$.

In the mass independent scheme the points $((w_0^e m_{\api})^2,w_0^e T_c)$
are linearly interpolated, and the deconfinement temperature is extrapolated
to
\begin{equation}
M^{w^e_0}_R \doteq (w_0^e m_{\api})^2=0. 
\end{equation}
The linear fit shown in Fig.~\ref{extrapolation-mass-independent} clearly
indicates that the deconfinement transition occurs at lower temperatures
when the gluino mass is decreased,
\begin{equation}
 w_0^e\, T_c(M^{w_0^e}_R) = 0.0190(22)M^{w^e_0}_R + 0.2432(45).
\end{equation}
The final extrapolation to the supersymmetric limit $M^{w_0^e}_R = 0$ leads
to
\begin{equation}
 w_0^e\, T_c = 0.2432(45),
\end{equation}
where the quoted error is only statistical. 

As an alternative we employ a mass-dependent renormalisation scheme. The
points $((w_0 m_{\api})^2,w_0 T_c)$ with $w_0$ determined at the same value
of the bare parameters $\kappa$ and $\beta$ at the phase transition are
linearly interpolated,
\begin{equation}\label{massdependent}
 w_0\, T_c(M^{w_0}_R) = 0.01234(7)M^{w_0}_R + 0.2441(26),
\end{equation}
where $M^{w_0}_R \doteq (w_0 m_{\api})^2$. The interpolation is shown in
Fig.~\ref{extrapolation-mass-dependent}. The slope is different due to the
change of renormalisation scheme. However, the final extrapolation to the
supersymmetric limit leads to the compatible result
\begin{equation}
 w_0 T_c = 0.2441(26),
\end{equation}
but with a smaller error due to the more precise determination of the scale
for points with heavier gluino masses.

For a comparison of $\mathcal{N} = 1$ SYM and pure gauge theory, we computed
the scale $w_0/a$ at infinite gluino mass on a lattice $18^4$ and $\beta=
1.829$. The chosen $\beta$ is the critical value for the deconfinement
transition of pure SU(2) Yang-Mills theory with a Symanzik improved gauge
action at $\Nt = 6$ \cite{Cella:1993ic}. The measured value of
\begin{equation}
w_0/a= 1.7649(78)
\end{equation}
leads to 
\begin{equation}
w_0 T_c = 0.2941(13)
\end{equation}
for pure SU(2) Yang-Mills theory.

The ratio of the deconfinement temperatures for pure and supersymmetric
Yang-Mills theory is thus
\begin{equation}
 \frac{T_c(\textrm{SYM})}{T_c(\textrm{pure Yang-Mills})} = 0.826(18).
\end{equation}
Introducing physical units by setting $T_c = 240$ MeV for the critical
temperature in pure gauge theory, we obtain a physical value of the
deconfinement phase transition temperature for $\mathcal{N}=1$ SU(2)
supersymmetric Yang-Mills theory,
\begin{equation}
T_c = 198(4)\ \textrm{MeV}.
\end{equation}
This value is consistent with the one level stout data, see
\Eqref{eq:roughstout1}, where we obtain the rough estimate
\begin{equation}
T_c = 227(32)\ \textrm{MeV}
\end{equation}
from our data, assuming $\rne=0.5\ \textrm{fm}$ as in QCD.

In the analysis of the deconfinement transition we have found no evidence
for contributions from negative Pfaffians even at the largest values of
$\kappa$.

\section{The chiral phase transition}

The order parameter of the chiral phase transition is the gluino condensate.
A non-zero expectation value of this parameter signals the breaking of the
$Z_2$ remnant of the U(1)$_R$ symmetry. The bare gluino condensate is
defined as the derivative of the logarithm of the partition function with
respect to the bare gluino mass parameter,
\begin{equation}
\langle \bar{\lambda} \lambda \rangle_B 
\doteq -\frac{T}{V}\frac{\partial}{\partial m} \log(Z(\beta,m)).
\end{equation}
Chiral symmetry is broken by our lattice action with the Wilson-Dirac
operator for the fermions, and the bare gluino condensate
$\langle\bar{\lambda} \lambda \rangle_B$ acquires an additive and
multiplicative renormalisation:
\begin{equation}
\label{condensate-ren}
\langle \bar{\lambda} \lambda \rangle_R 
= Z_{\bar{\lambda} \lambda}(\beta)
(\langle \bar{\lambda} \lambda \rangle_B - b_0).
\end{equation}

At zero temperature a first order transition is expected when the bare
gluino mass is changed, crossing a critical value corresponding to $M_R=0$.
Close to such a transition the histogram of $\langle \bar{\lambda} \lambda
\rangle$ shows a two peak structure in a finite volume. The transition can
be identified with the point where the symmetry of the two peaks changes, as
done in \cite{Kirchner:1998mp}. Such an analysis is independent of the
renormalisation described in \Eqref{condensate-ren}. At finite temperatures
the first order chiral phase transition extends to a phase transition line
at $M_R=0$, terminating in a second order end-point. Beyond that point the
transition changes from first order to a cross over, see
Fig.~\ref{scenarios}. For this reason, considerations on the renormalisation
procedure become important for the precise localisation of the phase
transition.

The additive renormalisation is removed by a subtraction of the zero
temperature result:\footnote{Notice that with this convention the chiral
condensate will be zero at zero temperature and non-zero at higher
temperatures.}
\begin{equation}
\langle \bar{\lambda} \lambda \rangle_S = \langle \bar{\lambda} \lambda \rangle_B^{T=0} - \langle \bar{\lambda} \lambda \rangle_B^{T}\; .
\end{equation}
The calculation of the renormalisation constant $Z_{\bar{\lambda}
\lambda}(\beta)$ can be avoided in a fixed scale approach, where the bare
coupling $\beta$ and $\kappa$ are fixed and the temperature is changed by a
variation of $\Nt$.

The bare gluino condensate is obtained from the trace of the inverse
Wilson-Dirac operator,
\begin{eqnarray}
-\frac{T}{V}\frac{\partial}{\partial m} \log(Z(\beta,m)) 
& = & -\frac{1}{Z(\beta,m)}\frac{T}{V}\frac{\partial}{\partial m} 
\left\langle \exp{ \left(\frac{1}{2}\tr\log(D_W(m)) \right)} 
\right\rangle_{S_g} \nonumber \\
& = & -\frac{T}{V} \left\langle \frac{1}{2} \tr (D_W^{-1}) \right\rangle.
\end{eqnarray}
Here and in the following $\langle O \rangle_{S_g}$ denotes the functional
integral with respect to the gauge part of the action, i.~e.\ $Z(\beta,m) =
\left\langle \exp{ \left(\frac{1}{2}\tr\log(D_W(m)) \right)}
\right\rangle_{S_g}$, where a positive Pfaffian is assumed. The trace of the
inverse Wilson-Dirac operator is evaluated with 20 random noise vectors
using the stochastic estimator technique.

The simulations are done on a lattice $12^3 \times \Nt$, with $\Nt \in
\{4,\dots,11\}$, $\beta=1.7$, and $\kappa=0.192$. A simulation at zero
temperature, i.~e.\ $\Nt=12$, has been performed to determine the adjoint
pion mass $am_{\api}=0.388(9)$ and the value of the scale $w_0/a=2.070(38)$.

\begin{figure}[tb]
\centering
\subfigure[Subtracted chiral condensate]
{\includegraphics[width=0.47\textwidth]{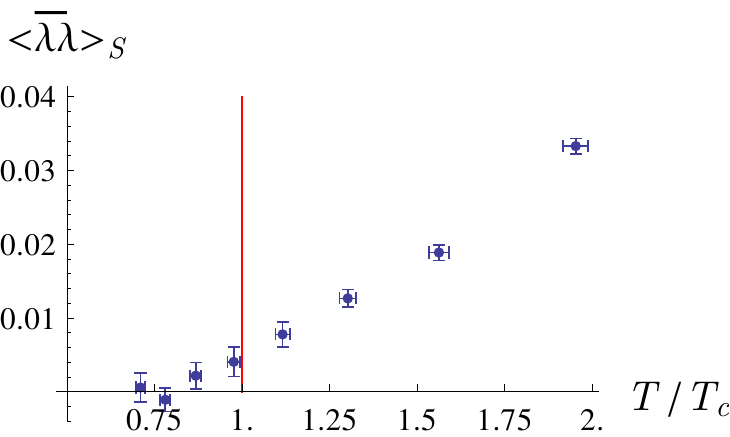}
\label{chiralcond12c}}
\subfigure[Chiral susceptibility]
{\includegraphics[width=0.47\textwidth]{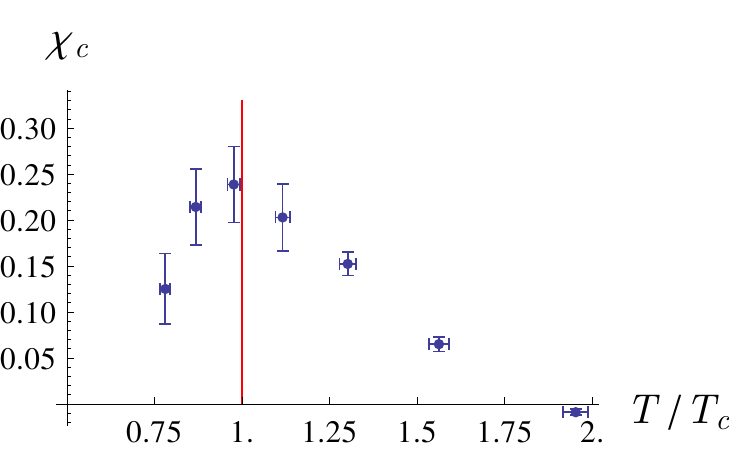}
\label{chiralcondsusc12c}}
\subfigure[Polyakov loop]
{\includegraphics[width=0.47\textwidth]{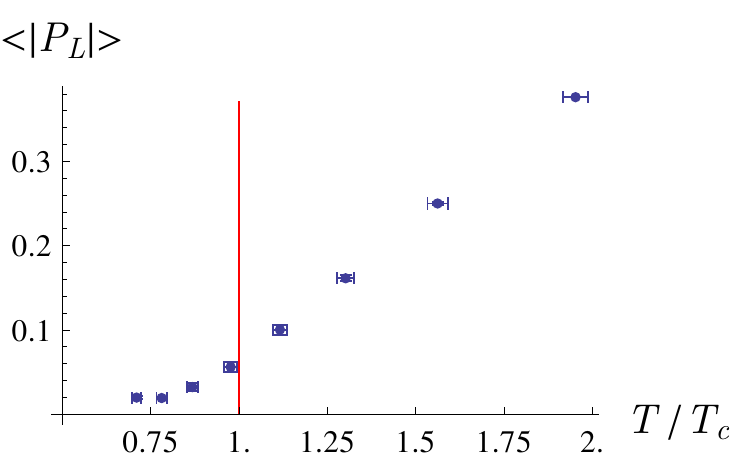}
\label{polyakov12c}}
\subfigure[Polyakov loop susceptibility]
{\includegraphics[width=0.47\textwidth]{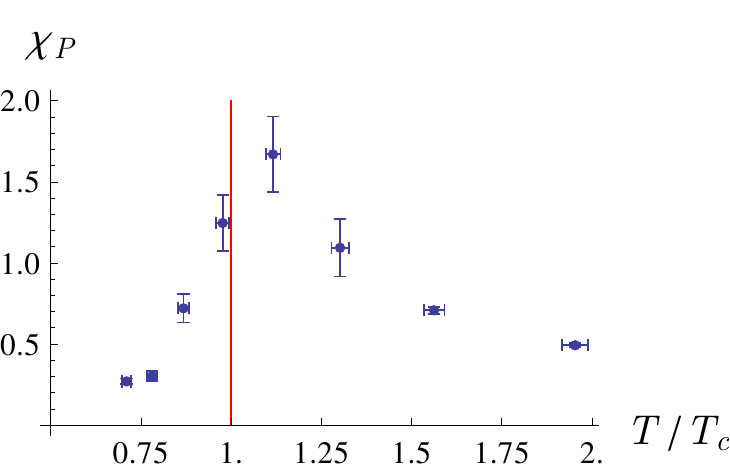}
\label{polyakovsusc12c}}
\caption{a-b) Chiral condensate and its susceptibility on a $12^3\times \Nt$
lattice at $\beta = 1.7$, $\kappa = 0.192$; c-d) Polyakov loop and its
susceptibility at the same parameters. $T_c$ refers to the deconfinement
transition temperature obtained from an extrapolation to the supersymmetric limit using
\Eqref{massdependent}.}
\end{figure}

\begin{figure}[tb]
\centering
\subfigure{
\includegraphics[width=0.47\textwidth]{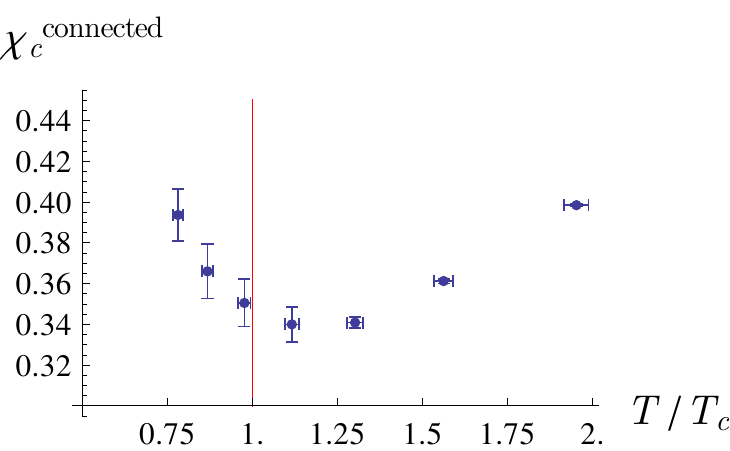}
\label{conchiralcondsusc12c}}
\subfigure{
\includegraphics[width=0.47\textwidth]{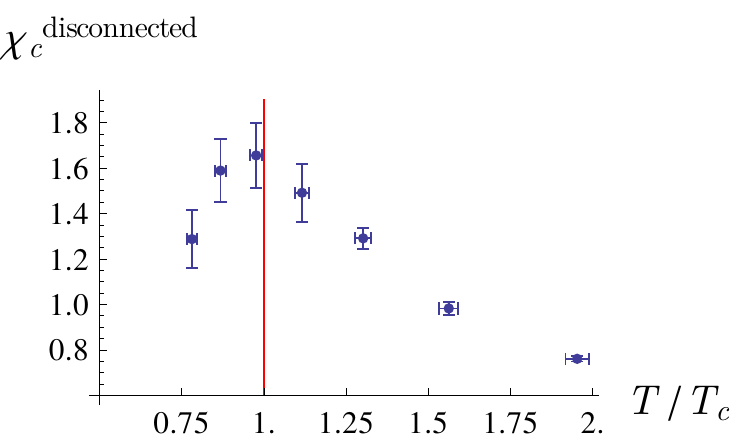}
\label{dischiralcondsusc12c}}
\caption{Comparison of connected (left) and disconnected (right) contributions to the chiral susceptibility on a $12^3\times \Nt$ lattice at $\beta = 1.7$, $\kappa = 0.192$.
\label{connecteddisconnecedchiraldistribution}}
\end{figure}

The subtracted chiral condensate starts to rise at $\Nt \simeq 7$, but its
behaviour is quite smooth due the crossover nature of the transition away
from the supersymmetric limit, see Fig.~\ref{chiralcond12c}. For a better
identification of the pseudo-critical transition point, we determine the
peak of the chiral susceptibility $\chi_c$. This observable is proportional
to the derivative of the gluino condensate and it has connected and
disconnected contributions:
\begin{eqnarray}
\chi_c 
& = & -\frac{T}{V} \frac{\partial^2}{\partial m^2} \log(Z(g,m)) 
= -\frac{T}{V} \frac{\partial}{\partial m} 
\left\langle \frac{1}{2} \tr (D_W^{-1}) \exp{ 
\left(\frac{1}{2}\tr\log(D_W(m)) \right)}  \right\rangle_{S_g} \nonumber \\
& = & -\frac{T}{V} \left\{\left\langle \frac{1}{4} 
\tr (D_W^{-1})^2 \right\rangle 
- \left\langle \frac{1}{4} \tr (D_W^{-1}) \right\rangle^2 
- \left\langle \frac{1}{2} \tr (D_W^{-2}) \right\rangle \right\}.
\end{eqnarray}
The connected contribution is expected to vanish in the supersymmetric
limit. In the range of parameters that we have considered in this
investigation the disconnected contribution is already dominant and the
connected contribution can be neglected for a localisation of the peak, see
Fig.~\ref{connecteddisconnecedchiraldistribution}. In that respect the
relevant dynamics of the phase transition is already similar to the one at a
vanishing residual gluino mass. Even though the connected contribution is
negligible, we consider the complete observable in the following. In that
way we ensure that absence of additional systematic uncertainties in our
extrapolations.


The results of our simulation are shown in Fig.~\ref{chiralcondsusc12c}.
Even though there is quite a broad central region, a visible peak can be
identified corresponding to the value at $\Nt = 9$. For comparison, the
Polyakov loop is shown in Fig.~\ref{polyakov12c}. It acquires a
non-vanishing expectation value at $\Nt = 8$. This small deviation is still
consistent with the scenario of a chiral symmetry restoration and a
deconfinement phase transition at the same temperature.

\begin{figure}[tb]
\centering
\subfigure[$\Nt = 5$]{
\includegraphics[width=0.47\textwidth]{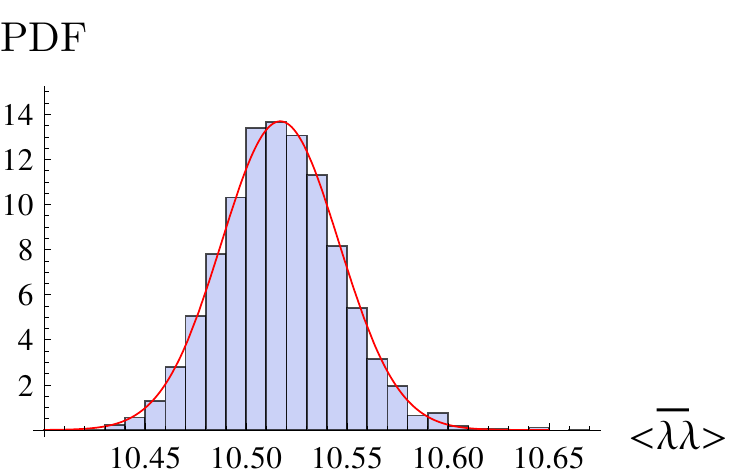}}
\subfigure[$\Nt = 6$]{
\includegraphics[width=0.47\textwidth]{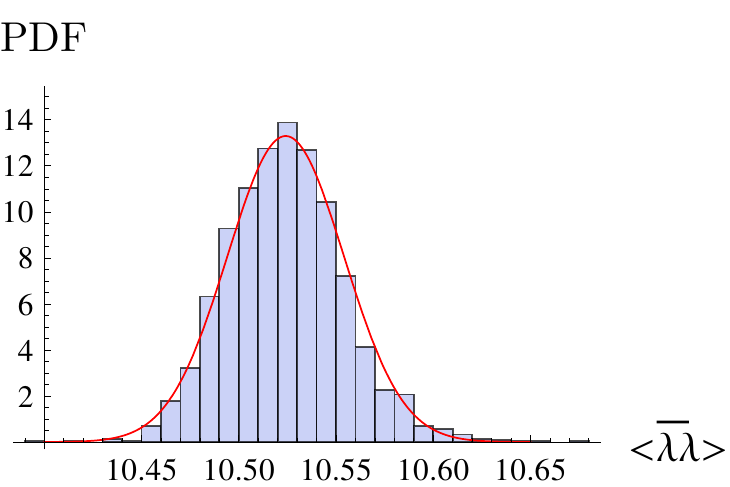}}
\subfigure[$\Nt = 7$]{
\includegraphics[width=0.47\textwidth]{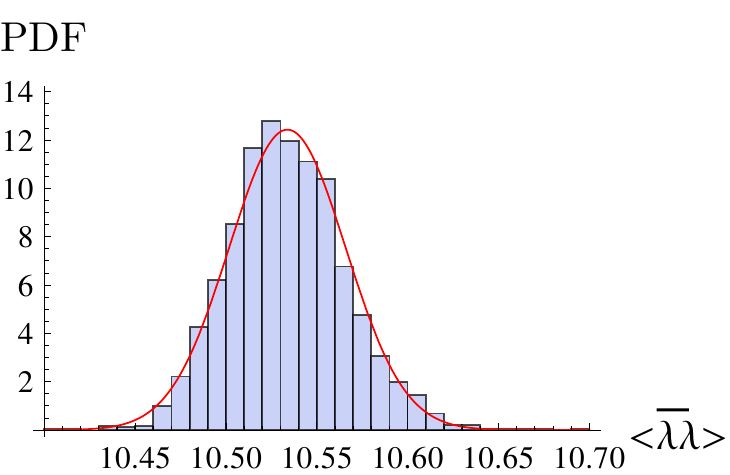}}
\subfigure[$\Nt = 8$]{
\includegraphics[width=0.47\textwidth]{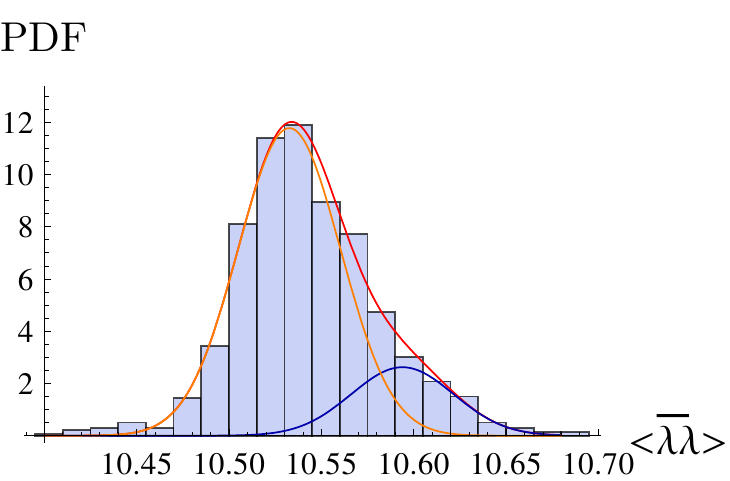}}
\caption{Distributions of the chiral condensate on $12^3\times \Nt$ lattices
at $\beta =1.7$ and $\kappa = 0.194$. For $\Nt = 5$ the distribution is
compatible with a single Gaussian, while for $\Nt = 8$ a second peak
emerges.
\label{chiraldistribution}}
\end{figure}

In order to provide an upper limit for the chiral symmetry restoration
temperature, we have done new simulations approximately at the
supersymmetric limit, i.~e.\ around the value of $\kappa$ where the adjoint
pion mass is expected to vanish, following the approach of
\cite{Kirchner:1998mp}. This corresponds to $\kappa = 0.194$ at $\beta
=1.7$. The lattice sizes were chosen to be $12^3\times \Nt$ with $\Nt \in
\{5,6,7,8\}$. Note that these simulations cannot be done at large values of
$\Nt$, due to the long time needed for a convergence of the conjugate
gradient algorithm in that limit. The distributions are displayed in
Fig.~\ref{chiraldistribution}. At high temperatures, like $\Nt = 5$, the
distribution is close to a Gaussian without any indications for a double
peak in the gluino condensate. At $\Nt=8$, on the other hand, we observe a
small second peak emerging from the distribution. Therefore, at these low
temperatures the transition, close to the supersymmetric limit, becomes
consistent with a first order chiral phase transition. Moreover, this
suggests that the transition happens in the region between $\Nt=7$ and
$\Nt=8$. It is another indication that the chiral phase transition and the
deconfinement transition are close to each other: assuming that the value of
$w_0$ does not change so much going from $\kappa=0.192$ to $\kappa=0.194$ we
are able to estimate an upper limit for the supersymmetric chiral critical
temperature: $T_{\chi}(M_R=0) \lesssim 1.5\, T_c$.

We have checked our assumption of a positive Pfaffian by a measurement of
its sign on 200 configurations. At $\kappa=0.192$ we have found no
contribution, whereas at $\kappa=0.194$ around $8\%$ of the configurations
have a negative sign on the $12^3\times8$ lattice. Hence the simulations at
the supersymmetric limit provide only an estimate of the transition point.
In further studies the contributions with a negative sign have to be taken
into account more carefully; alternatively, with a larger amount of
computing time, the supersymmetric limit can extrapolated from a region
without a relevant sign problem.

\section{Conclusions}

We have investigated the deconfinement and the chiral phase transitions in
$\mathcal{N}=1$ supersymmetric Yang-Mills theory. Different from the case of
QCD, the Polyakov loop is a well-defined order parameter at all values of
the gluino mass, and the deconfinement transition can be identified in an
unambiguous way. The U(1)$_R$ chiral symmetry is only partially broken by
the anomaly and a remnant $Z_{2N_c}$ symmetry survives. The expectation
value of the gluino condensate is the order parameter for the breaking of
this remnant symmetry down to $Z_2$.

We have investigated the dependence of both order parameters on the
temperature and on the gluino mass. We have determined the temperature,
where the deconfinement phase transition takes place, with a good accuracy,
and extrapolated the transition temperature in the supersymmetric limit.
Different scale setting prescriptions lead to consistent results, indicating
the reliability of the result. The transition happens at a temperature,
which is around 80\% of the transition temperature in pure Yang-Mills
theory.

The identification of the chiral phase transition point, on the other hand,
needs more effort, since the transition becomes a crossover at finite gluino
masses. In a fixed scale approach we have identified the transition region
taking also the renormalisation into account. We were able to narrow the
range for the chiral transition down to a region close to the deconfinement
transition. This situation can be compared with $N_f = 2$ adjoint QCD
(aQCD), a theory similar to SYM. In the case of aQCD, there exists a mixed
phase with deconfinement but a broken chiral symmetry. The deconfinement
temperature is eight times smaller than the point of chiral symmetry
restoration \cite{Karsch:1998qj}. From this perspective, SYM appears to be
more similar to QCD, where the deconfinement and chiral symmetry restoration
seem to occur at the same temperature.

In order to confirm scenario (a) of Fig.~\ref{scenarios} with coincident
phase transitions, it will be necessary to perform simulations in more
parameter points, with higher statistics, and on larger lattices. A study of
the finite size scaling is of great importance to test the existence of a
second order endpoint for the chiral phase transition in the supersymmetric
limit.

Presently we have been able to study the phase transitions only at rather
low values of $\beta$, i.~e.\ at relatively large lattice spacings. At these
parameters there is still a considerable deviation from the degeneracy of
the particle masses in supermultiplets. Hence studies at larger values of
$\Nt$ are needed for reliable extrapolations to the supersymmetric limit.
This is the largest source of a systematic uncertainty for our current
determination of the deconfinement transition point.

In the future we also plan to study different numbers of colours $N_c$, and
different boundary conditions, since the phase transitions seem to be
sensitive to these parameters.

\section*{Acknowledgements}

We thank I.~Montvay for helpful instructions and comments. The authors
gratefully acknowledge the computing time granted by the John von Neumann
Institute for Computing (NIC) and provided on the supercomputer JUROPA at
J\"ulich Supercomputing Centre (JSC), and by the Leibniz-Rechenzentrum (LRZ)
in M\"unchen provided on the supercomputer SuperMUC. Further computing time
has been provided by the compute cluster PALMA of the University of
M\"unster.


\newpage
\begin{appendix}
 \numberwithin{table}{section}

\section{Details of the simulations}
\begin{table}[ht]
\centering
\begin{small}
\begin{tabular}{ccllclr}
\hline\hline
$\Nt$ & $N_s$ & \hspace{3mm}$\beta$ & \hspace{3mm}$\kappa$ & $r_0/a$ 
& \hspace{3mm}$am_{\api}$ & $\Ncon$ \\
\hline
32 & 16 & 1.65 & 0.1150 & -- & 2.206(14) & 600 \\
32 & 16 & 1.55 & 0.1475 & -- & 1.2770(24) & 621 \\
32 & 16 & 1.5 & 0.155 & -- & 1.1316(34) & 800 \\
32 & 16 & 1.5 & 0.158 & 2.68(6) & 0.97863(84) & 2036 \\
32 & 16 & 1.5 & 0.160 & 2.85(3) & 0.8570(20)  & 2177 \\
32 & 16 & 1.5 & 0.162 & 3.11(9) & 0.7085(19)  & 2076 \\
32 & 16 & 1.5 & 0.163 & 3.15(8) & 0.6199(12)  & 2001 \\
32 & 16 & 1.5 & 0.164 & 3.34(8) & 0.5066(26) & 1720 \\
24 & 12 & 1.45 & 0.1625 & -- & 0.9865(25) & 1000 \\
24 & 12 & 1.40 & 0.145 & -- & 1.722(40) & 1199 \\
24 & 12 & 1.40 & 0.150 & -- & 1.563(72) & 1599 \\
24 & 12 & 1.40 & 0.153 & -- & 1.501(13) & 1800 \\
24 & 12 & 1.40 & 0.155 & -- & 1.434(18) & 1999 \\
24 & 12 & 1.40 & 0.160 & -- & 1.2858(13) & 1240 \\
\hline
\end{tabular}
\end{small}
\caption{Parameters for the additional zero temperature simulations with one
level of stout smearing. At $\beta=1.5$ the value of $r_0/a$ is extrapolated
to the supersymmetric limit.}
\label{tablemeasstout2}
\end{table}
\begin{table}[ht]
\centering
\begin{small}
\begin{tabular}{l|ccccc}
\hline
$\beta$   &1.5  & 1.6     & 1.75     & 1.9        & 2.1      \\
\hline  
$\rne/a$  & 3.81(12) & 5.93(5) & 9.02(18) &  12.20(12) & 16.56(39) \\
\hline
\end{tabular}
\end{small}
\caption{Values of $r_0^e/a$ determined for different values of $\beta$. The
values for $\beta=1.6$ and $1.75$ can be found in our previous publications;
the others are presented here for the first time. The value at $\beta=1.5$
is determined from the extrapolation of the zero temperature results in
Table \ref{tablemeasstout2}.}
\label{betar0values}
\end{table}
\begin{table}[ht]
\centering
\begin{small}
\begin{tabular}{crcrl}
\hline\hline
$\Nt$ & $N_s$ & $\beta$ & $\Ncon$ & \hspace{5mm}$\kcd$ \\
\hline
4 & 12 &  1.60 & 2500  &0.140(15) \\
4 & 12 &  1.50 & 2500  &0.155(10) \\
4 &  8 &  1.70 & 20000 &0.0000(25) \\
4 &  8 &  1.65 & 20000 &0.1150(50) \\
4 &  8 &  1.60 & 20000 &0.1350(50) \\
4 &  8 &  1.55 & 20000 &0.1475(50) \\
4 &  8 &  1.50 & 20000 &0.1550(50) \\
4 &  8 &  1.45 & 20000 &0.1625(25) \\
4 &  8 &  1.40 & 20000 &0.1650(25) \\
\hline
\end{tabular}
\end{small}
\caption{The number of measured configurations $\Ncon$, produced at the
$\kcd$ value, used for estimating the deconfinement transition, using one
level of stout smearing. The parameters are the same as in our zero
temperature investigations of the particle spectrum \cite{Bergner:2013nwa}.}
\label{tablemeasstout}
\end{table}
\begin{table}[ht]
\centering
\begin{small}
\begin{tabular}{ccccccc}
\hline\hline
$\Nt$ & $N_s$ & $\beta$ & $\kappa$ & $a m_{\api}$ &  $w_0/a$ & $w_0^e/a$ \\
\hline
20 & 10  &  1.65 & 0.1600 & 1.7182(09) & 1.179(02) & 1.428(36) \\
20 & 10  &  1.65 & 0.1825 & 1.1138(20) & 1.310(09) & 1.428(36) \\
20 & 10  &  1.65 & 0.1850 & 1.0277(24) & 1.340(09) & 1.428(36) \\
20 & 10  &  1.65 & 0.1875 & 0.9342(22) & 1.326(13) & 1.428(36) \\
20 & 10  &  1.62 & 0.1900 & 0.9398(26) & 1.258(23) & 1.359(34) \\
20 & 10  &  1.62 & 0.1925 & 0.8331(29) & 1.297(23) & 1.359(34) \\
20 & 10  &  1.62 & 0.1950 & 0.7067(44) & 1.384(31) & 1.359(34) \\
20 & 10  &  1.60 & 0.1950 & 0.8159(62) & 1.277(20) & 1.307(30) \\
20 & 10  &  1.60 & 0.1975 & 0.6868(40) & 1.351(26) & 1.307(30) \\
20 & 10  &  1.60 & 0.2000 & 0.4980(58) & 1.553(31) & 1.307(30) \\
\hline
\end{tabular}
\end{small}
\caption{The table summarises the zero temperature measurements done for
setting the scale without stout smearing. $w_0/a$ is the mass dependent
value while $w_0^e/a$ is the mass independent one, i.~e.\ obtained by
extrapolation to $(w_0 m_{\api})^2 = 1$.}
\label{tablemeaszero}
\end{table}
\begin{table}[ht]
\centering
\begin{small}
\begin{tabular}{cccrcc}
\hline\hline
$\Nt$ & $N_s$ & $\beta$ & $\Ncon$ & $\tau$ & $\kcd$ \\
\hline
4 & 8  &  1.65 & 150000 & 400  & 0.1600(50) \\
4 & 12 &  1.65 & 80000  & 1100 & 0.1600(50) \\
4 & 16 &  1.65 & 40000  & 1600 & 0.1580(50) \\
5 & 15 &  1.65 & 20000  & 1500 & 0.1850(25) \\
5 & 15 &  1.62 & 20000  & 1500 & 0.1925(20) \\
5 & 15 &  1.60 & 20000  & 1500 & 0.1950(20) \\
\hline
\end{tabular}
\end{small}
\caption{The table summarises the number of measurements $\Ncon$, produced
at the $\kcd$ value, for estimating the deconfinement transition using the
action without stout smearing. The autocorrelation time $\tau$ is computed
for the Polyakov loop at the critical point.}
\label{tablemeasunstout}
\end{table}
\end{appendix}

\end{document}